\documentclass{aa}

\usepackage{amsmath}
\usepackage{graphicx}
\usepackage{graphics}
\usepackage{epsfig}
\usepackage{longtable}
\usepackage{blindtext}
\usepackage{txfonts}
\usepackage[colorlinks=true,linkcolor=blue,citecolor=blue,urlcolor=blue]{hyperref}%
\begin{document}

\title{Cold gas in the early Universe}
\subtitle{Survey for neutral atomic-carbon in GRB host galaxies at $1 < z < 6$ from \\ optical afterglow spectroscopy}
\titlerunning{C\,\textsc{i} absorption in GRB afterglows}

\author{
K.~E.~Heintz\inst{1,2,3},
C. Ledoux\inst{4},
J. P. U. Fynbo\inst{2,3},
P. Jakobsson\inst{1},
P.~Noterdaeme\inst{5},
J.-K.~Krogager\inst{5},
J. Bolmer\inst{4}, 
P. M\o ller\inst{6}, 
S.~D.~Vergani\inst{7,5}, 
D. Watson\inst{2,3},
T.~Zafar\inst{8},
A.~De~Cia\inst{6},
N.~R.~Tanvir\inst{9},
D. B. Malesani\inst{2,3,10}, 
J.~Japelj\inst{11}, 
S.~Covino\inst{12}, \&
L.~Kaper\inst{11}
}
\institute{
Centre for Astrophysics and Cosmology, Science Institute, University of Iceland, Dunhagi 5, 107 Reykjav\'ik, Iceland
\and
Cosmic Dawn Center, Niels Bohr Institute, University of Copenhagen, Juliane Maries Vej 30, 2100 Copenhagen \O, Denmark
\and
 DTU-Space, Technical University of Denmark, Elektrovej 327, 2800 Kgs.\ Lyngby, Denmark \\
\email{keh14@hi.is}
\and
European Southern Observatory, Alonso de C\'ordova 3107, Vitacura, Casilla 19001, Santiago 19, Chile
\and
Institut d'Astrophysique de Paris, CNRS-UPMC, UMR7095, 98bis bd Arago, 75014 Paris, France
\and
European Southern Observatory, Karl-Schwarzschildstrasse 2, D-85748 Garching bei M\"unchen, Germany
\and
GEPI, Observatoire de Paris, PSL Research University, CNRS, Place Jules Janssen, 92190 Meudon
\and
Australian Astronomical Observatory, PO Box 915, North Ryde, NSW 1670, Australia
\and
Department of Physics \& Astronomy and Leicester Institute of Space \& Earth Observation, University of Leicester, University Road, Leicester LE1 7RH, United Kingdom
\and
Dark Cosmology Centre, Niels Bohr Institute, University of Copenhagen, Juliane Maries Vej 30, 2100 Copenhagen \O, Denmark
\and
Anton Pannekoek Institute for Astronomy, University of Amsterdam, Science Park 904, 1098 XH Amsterdam, The Netherlands
\and
INAF -- Osservatorio Astronomico di Brera, Via E. Bianchi 46, I-23807 Merate (LC), Italy 
}
\authorrunning{Heintz et al.}

\date{Received 2018}

\abstract{
We present a survey for neutral atomic-carbon (C\,\textsc{i}) along gamma-ray burst (GRB) sightlines, which probes the shielded neutral gas-phase in the interstellar medium (ISM) of GRB host galaxies at high redshift. We compile a sample of 29 medium- to high-resolution GRB optical afterglow spectra spanning a redshift range through most of cosmic time from $1 < z < 6$. We find that seven ($\approx 25\%$) of the GRBs entering our statistical sample have C\,\textsc{i} detected in absorption. It is evident that there is a  strong excess of cold gas in GRB hosts compared to absorbers in quasar sightlines. We investigate the dust properties of the GRB C\,\textsc{i} absorbers and find that the amount of neutral carbon is positively correlated with the visual extinction, $A_V$, and the strength of the 2175\,\AA~dust extinction feature, $A_{\mathrm{bump}}$. GRBs with C\,\textsc{i} detected in absorption are all observed above a certain threshold of $\log N$(H\,\textsc{i})$/\mathrm{cm}^{-2}$ + [X/H] > 20.7 and a dust-phase iron column density of $\log N$(Fe)$_{\mathrm{dust}}/\mathrm{cm}^{-2}$ > 16.2. In contrast to the SED-derived dust properties, the strength of the C\,\textsc{i} absorption does not correlate with the depletion-derived dust properties. This indicates that the GRB C\,\textsc{i} absorbers trace dusty systems where the dust composition is dominated by carbon-rich dust grains. The observed higher metal and dust column densities of the GRB C\,\textsc{i} absorbers compared to H$_2$- and C\,\textsc{i}-bearing quasar absorbers is mainly a consequence of how the two absorber populations are selected, but is also required in the presence of intense UV radiation fields in actively star-forming galaxies. 
}

\maketitle

%%%%%%%%%%%%%%%%%%%%%%%%%%%%%%%%%%%%%%%%%%%%%%%%%%%%%%%%%%%%%%%%%%%%%%%%%%%%
\section{Introduction}     
\label{sec:introduction}
%%%%%%%%%%%%%%%%%%%%%%%%%%%%%%%%%%%%%%%%%%%%%%%%%%%%%%%%%%%%%%%%%%%%%%%%%%%%

Long-duration gamma-ray bursts (GRBs) are the most energetic class of cosmic explosions and are believed to originate from the deaths of massive stars \citep[e.g.][]{Woosley06}. Their immense brightness makes them detectable even out to the epoch of reionization at $z\gtrsim 8$ \citep[][]{Salvaterra09,Tanvir09,Tanvir18,Cucchiara11}. Since GRBs are expected to probe star formation \citep{Wijers98,Christensen04,Jakobsson05,Kistler09,Robertson12,Greiner15} and star-formation rates are driven by the availability of dense gas, GRBs offer a potentially ideal probe of the physical conditions in the cold, neutral gas-phase of the interstellar medium (ISM) in star-forming galaxies through most of cosmic time.

It is even possible to detect absorption by molecular hydrogen in the host directly by the presence of Lyman-Werner bands. However, to date, only four GRB afterglow spectra have a robust detection of H$_2$ in absorption \citep[][]{Prochaska09,Delia14,Kruehler13,Friis15}. One explanation for the dearth of molecular hydrogen in GRB host galaxies could be that the H$_2$ molecules are photodissociated by the GRB event itself. The intense prompt $\gamma$-ray flash and afterglow emission will, however, only impact gas in the vicinity of the GRB \citep[out to $\lesssim 10$ pc;][]{Draine02}. The gas producing absorption lines in GRB optical spectra is typically found to be located several hundred parsecs from the GRB explosition site \citep{Delia07,Prochaska06,Vreeswijk07,Vreeswijk11}. If there is a lack of detection of cold and molecular gas, it should be intrinsic to the GRB host galaxy properties.

While quasar selection is subject to a significant bias excluding sightlines with intervening dusty and therefore also metal-rich foreground galaxies \citep[][Krogager et al. subm.]{Fall93,Boisse98,Vladilo05,Pontzen09,Fynbo13,Fynbo17,Krogager15,Krogager16b,Krogager16a,Heintz18b}, the high-energy emission from GRBs on the other hand is unaffected by dust. The spectroscopic follow-up observations of GRB afterglows are, however, still much less likely to be successful for the most dust-obscured sightlines, the so-called "dark" bursts \citep{Fynbo01,Fynbo09,Jakobsson04,VanDerHorst09,Perley09,Perley13,Greiner11,Kruhler11,Watson12} since observations with high-resolution spectrographs and good signal-to-noise ratios (due to the transient nature of GRBs) are only obtained for the brightest afterglows \citep{Ledoux09}. 

It is still unclear whether the apparent lack of molecular gas in GRB host galaxies is an intrinsic property or simply due to observational limitations (though an effort has recently been made by \citealt{Bolmer18b} to quantify this). The absorption signatures of H$_2$ are the Lyman and Werner bands which are located bluewards of the Lyman-$\alpha$ (Ly$\alpha$) absorption line and in the Ly$\alpha$ forest. High-resolution spectra are therefore helpful to disentangle the H$_2$ transitions from the Ly$\alpha$ forest and to derive reliable H$_2$ column densities. As stated above, such samples will suffer from a bias against the most metal-rich and dust-obscured bursts \citep{Ledoux09}. These host galaxy properties are, however, found to be essential for the detection probability of H$_2$ \citep{Ledoux03,Petitjean06,Noterdaeme08,Noterdaeme15,Kruehler13}. In addition, due to the UV-steep extinction curves typically observed in GRB sightlines \citep{Zafar18a}, the observed flux in the Lyman-Werner bands will be attenuated much more strongly than the visual extinction, $A_V$. The sightlines most likely to exhibit H$_2$ absorption are therefore also the most difficult to observe due to the faintness of the afterglows.

The absorption signatures from neutral atomic-carbon (C\,\textsc{i}) are another probe of the shielded gas-phase and can be used to study the cold, neutral medium in the ISM. \cite{Ledoux15} searched for the absorption features of C\,\textsc{i} in a large sample of quasar spectra from the Sloan Digital Sky Survey (SDSS). 
Neutral carbon is typically observed to be coincident with H$_2$ in quasar DLAs \citep{Srianand05} and is found to be the optimal tracer of molecular gas in low-metallicity, star-forming galaxies \citep{Glover16} based on numerical simulations. This is likely related to the ionization potential of C\,\textsc{i} (11.26~eV) being similar to the energy range of Lyman-Werner photons that can photodissociate H$_2$ (11.2 -- 13.6~eV). The C\,\textsc{i}\,$\lambda\lambda$\,1560,1656 transitions are also located far from the Ly$\alpha$ forest and can be identified even in low- to medium-resolution spectroscopy. The feasibility of this approach was verified by the spectroscopic follow-up campaign of the C\,\textsc{i}-selected quasar absorbers, where H$_2$ was detected in absorption in all cases \citep{Noterdaeme18}. This sample also shows a constant C\,\textsc{i} to CO ratio, suggesting that neutral carbon actually probes a deeper, more shielded regime of the ISM than the diffuse atomic and molecular gas-phases \citep[see also e.g.][]{Snow06}. 

In this paper, we follow a similar approach to \cite{Ledoux15} but survey C\,\textsc{i} absorption in GRB optical afterglow spectra. We compile a list of $z > 1$ GRB afterglows, spanning over a decade of optical and near-infrared (NIR) spectroscopic observations, by combining the sample presented by \cite{Fynbo09} and the GRB afterglow legacy survey by \cite{Selsing18}. Our goal is to investigate the environments of GRB hosts with neutral carbon detected in absorption (henceforth simply referred to as GRB C\,\textsc{i} absorbers or C\,\textsc{i} systems) in terms of their dust properties and chemical abundances. Since it is only possible to observe the Ly$\alpha$ transition from the ground and thus derive gas-phase abundances at $z\gtrsim 1.7$, we rely on the GRBs in this redshift range to examine the chemical abundances of the GRB C\,\textsc{i} absorbers, whereas the dust properties can be studied at all redshifts. This survey also opens a potential route to test whether the proposed lack of molecular gas in GRB host galaxies is an intrinsic property or simply a consequence of observational limitations. In this paper, we present the results of this survey and the basic properties of GRB C\,\textsc{i} absorbers. These are compared to GRB host absorption systems without the presence of C\,\textsc{i} and to other types of "cold gas" absorbers (i.e. with H$_2$ and/or C\,\textsc{i} detected in absorption) in quasar sightlines.

The paper is structured as follows. In Sect.~\ref{sec:sel} we present our sample and describe the observations and selection criteria, where Sect.~\ref{sec:met} is dedicated to discuss the sample properties and completeness. In Sect.~\ref{sec:results} we show our results, with specific focus on the properties of GRB host galaxies with C\,\textsc{i} absorption compared to those without. For our discussion, we compare the sample of GRBs with prominent amounts of cold, neutral gas to similar samples of cold gas absorbers in quasar sightlines in Sect.~\ref{sec:disc}. Finally, in Sect.~\ref{sec:conc} we conclude on our work.

Throughout the paper, errors denote the $1\sigma$ confidence level and magnitudes are reported in the AB system. We assume a standard flat cosmology with $H_0 = 67.8$\,km\,s$^{-1}$\,Mpc$^{-1}$, $\Omega_m = 0.308$ and $\Omega_{\Lambda}=0.692$ \citep{Planck16}. Gas-phase abundances are expressed relative to the Solar abundance values from \cite{Asplund09}, where [X/Y] = $\log N(\mathrm{X})/N(\mathrm{Y}) - \log N(\mathrm{X})_{\odot}/N(\mathrm{Y})_{\odot}$.

%%%%%%%%%%%%%%%%%%%%%%%%%%%%%%%%%%%%%%%%%%%%%%%%%%%%%%%%%%%%%%%%%%%%%%%%%%%%
\section{The GRB sample}    \label{sec:sel}
%%%%%%%%%%%%%%%%%%%%%%%%%%%%%%%%%%%%%%%%%%%%%%%%%%%%%%%%%%%%%%%%%%%%%%%%%%%%

The sample of GRBs studied here was built by extracting all bursts from the VLT/X-shooter GRB (XS-GRB) afterglow legacy survey \citep{Selsing18}, combined with the GRB afterglow sample of \citet[][hereafter F09]{Fynbo09}. We also include the recent discovery of a strong C\,\textsc{i} absorber towards GRB\,180325A, which also shows a prominent 2175\,\AA~dust extinction feature \citep{Zafar18b}. We subsequently imposed additional observational selection criteria to exclude late-time host galaxy observations and afterglow spectra with poor spectral quality or low spectral resolution. In this section we detail the GRB afterglow observations (Sect.~\ref{ssec:obs}) and selection criteria (Sect.~\ref{ssec:sel}). In Sect.~\ref{sec:met} we present the properties of our final sample in terms of C\,\textsc{i} rest-frame equivalent width (Sect.~\ref{ssec:ew}) and dust extinction, $A_V$ (Sect.~\ref{ssec:av}). An overview of the full sample is provided in Table~\ref{tab:obs}.

\begin{table*}[!ht]
	\centering
	\begin{minipage}{\textwidth}
		\centering
		\caption{GRB afterglow sample properties.}
		\begin{tabular}{lcccccccl}
			\noalign{\smallskip} \hline \hline \noalign{\smallskip}
			GRB & $z_{\mathrm{GRB}}$ & $\log N(\textsc{H\,i})$ & $W_{\mathrm{r}}\,(\lambda\,1560)$ & $W_{\mathrm{r}}\,(\lambda\,1656)$ &  [X/H] & Ion & $A_V$ & Refs. \\
			& & (cm$^{-2}$)  & (\AA) & (\AA) & &&  (mag) & \\
			\noalign {\smallskip} \hline \noalign{\smallskip}
			\noalign {\smallskip} \noalign {\smallskip}
			F09 sample & &&&&&&& \\
			\noalign {\smallskip} \hline \noalign{\smallskip}
			050730 & 3.9693 & $22.10\pm 0.10$ & $< 0.03$ & $< 0.03$ & $-2.18\pm 0.11$  & S & $0.12\pm 0.02$ & (1, 2) \\
			050820A & 2.6147 & $21.05\pm 0.10$ & $< 0.12$ & $< 0.13$ & $-0.39\pm 0.10$  & Zn & $0.27\pm 0.04$ & (3, 4) \\
			050922C & 2.1995 & $21.55\pm 0.10$ & $< 0.03$ & $< 0.03$ & $-1.82\pm 0.11$ &  S & $0.09\pm 0.03$ & (3, 5) \\
			060210 & 3.9133 & $21.55\pm 0.15$ & $0.78\pm 0.03$ & $\cdots$ & $>-0.83$ & Si & $0.57\pm 0.08$ & (6, 5) \\
			060607A & 3.0749 & $16.95\pm 0.03$ & $< 0.05$ & $< 0.07$ & $\cdots$ &  $\cdots$ & $0.08\pm 0.08$ & (7, 8) \\
			061121 & 1.3145 & $\cdots$ & $0.49\pm 0.04$ &  $0.25\pm 0.04$ & $\cdots$ & $\cdots$  & $0.55\pm 0.10$ & (5) \\
			070802$^{\dagger}$ & 2.4511 & $21.50\pm 0.20$ & $0.70\pm 0.35$ & $1.36\pm 0.32$ & $-0.46\pm 0.63$ & Zn & $1.19\pm 0.15$ & (9, 2) \\
			071031 & 2.6918 & $22.15\pm 0.05$ & $< 0.03$ & $< 0.03$ & $-1.73\pm 0.05$ & Zn  & $<0.07$ & (3, 2) \\
			080310 & 2.4274 & $18.70\pm 0.10$ & $< 0.11$ & $< 0.14$ & $-1.20\pm 0.20$ & Si & $0.19\pm 0.05$ & (10, 8) \\
			080413A & 2.4330 & $21.85\pm 0.15$ & $< 0.03$ & $< 0.03$ & $-1.60\pm 0.16$ & Zn & $<0.59$ & (3, 11) \\
			080605$^{\dagger}$ & 1.6403 & $\cdots$ & $0.64\pm 0.23$ &  $0.38\pm 0.15$ & $\cdots$ &  $\cdots$ & $0.50\pm 0.13$ & (12)  \\
			080607$^{\star,\dagger}$ & 3.0368 & $22.70\pm 0.15$ & $2.17\pm 0.08$ & $2.03\pm 0.04$ & $>-0.20$ & O & $2.33\pm 0.46$ & (13, 2) \\
			\noalign {\smallskip} \hline \noalign{\smallskip}
			\noalign {\smallskip}\noalign {\smallskip}
			X-shooter sample &&&&&&&&\\
			\noalign {\smallskip} \hline \noalign{\smallskip}
			090926A & 2.1069  & $21.60\pm 0.07$ & $<0.05$ & $<0.06$ & $-1.85\pm 0.10$ & S & $<0.04$ & (14, 15) \\
			100814A & 1.4390 & $\cdots$ & $<0.06$ &  $<0.06$ & $\cdots$ & $\cdots$ & $<0.07$ & (15) \\
			120119A & 1.7288 & $22.44\pm 0.12$ & $0.51\pm 0.05$ &  $0.77\pm 0.06$ & $-0.96\pm 0.28$ & Zn  & $1.02\pm 0.11$ & (16, 15) \\
			120327A$^{\star}$ & 2.8148 & $22.01\pm 0.09$ & $<0.03$ & $<0.03$ & $-1.17\pm 0.11$ & Zn & $<0.03$ & (17) \\
			120815A$^{\star}$ & 2.3581 & $21.95\pm 0.10$ & ($0.12\pm 0.08$)\tablefootmark{$*$} & ($0.21\pm 0.05$)\tablefootmark{$*$} & $-1.15\pm 0.12$ & Zn & $0.19\pm 0.04$ & (18, 15) \\
			121024A$^{\star}$ & 2.3024 &  $21.88\pm 0.10$ & ($0.08\pm 0.05$)\tablefootmark{$*$} & ($0.11\pm 0.07$)\tablefootmark{$*$} & $-0.70\pm 0.10$ & Zn & $0.26\pm 0.07$  & (19, 15) \\
			130408A & 3.7579 & $21.70\pm 0.10$ & $<0.05$ & $<0.06$ & $-1.24\pm 0.12$ & S & $0.21\pm 0.05$ & (6, 20) \\
			130606A & 5.9129 & $19.91\pm 0.02$ & $<0.10$ & $<0.10$ & $-1.30\pm 0.08$ & Si  & $<0.07$ & (21, 15) \\
			141028A & 2.3334 & $20.55\pm 0.07$ & $<0.14$ & $<0.21$ & $-0.73\pm 0.34$ & Zn & $0.13\pm 0.09$ & (16) \\
			141109A & 2.9944 & $22.10\pm 0.20$ & $<0.11$ & $<0.11$ & $-1.40\pm 0.22$ & Zn & $0.11\pm 0.03^{\ddagger}$ & (22, 23) \\
			150403A & 2.0571 & $21.80\pm 0.20$ & $0.34\pm 0.03$ & $0.50\pm 0.04$ & $-0.80\pm 0.35$ & S & $0.12\pm 0.02^{\ddagger}$ & (22, 23) \\
			151021A & 2.3299 & $22.20\pm 0.20$ & $<0.17$ & $<0.28$ & $-1.11\pm 0.20$ & Si & $0.20\pm 0.03^{\ddagger}$  & (22, 23) \\
			151027B & 4.0647 & $20.50\pm 0.20$ & $<0.09$ & $<0.09$ & $-1.62\pm 0.24$ & Si & $0.10\pm 0.05$ & (22, 23, 20) \\
			160203A & 3.5185 & $21.75\pm 0.10$ & $<0.03$ & $<0.03$ & $-1.26\pm 0.11$ & S & $<0.10^{\ddagger}$ & (24) \\
			161023A & 2.7106 & $20.97\pm 0.01$ & $<0.03$ & $<0.03$ & $-1.11\pm 0.07$ & Zn & $0.09\pm 0.03$ & (25) \\
			170202A & 3.6450 & $21.55\pm 0.10$ & $<0.10$ & $<0.08$ & $-1.15\pm 0.13$ & S & $<0.12$ & (22, 23, 20) \\
			\noalign{\smallskip} \hline \noalign{\smallskip}
			180325A$^{\dagger}$ & 2.2486 & $22.30\pm 0.14$ & $0.58\pm0.05$ & $0.85\pm0.05$ & $>-0.98$ & Zn & $1.58\pm 0.12$ & (26) \\
			\noalign{\smallskip} \hline \noalign{\smallskip}
		\end{tabular}
		\tablefoot{All GRB afterglows with spectral coverage of the C\,\textsc{i}\,$\lambda\lambda$\,1560,1656 line transitions listed here are selected within a completeness limit of $W_{\mathrm{r}}(\lambda\,1560) > 0.2$\,\AA. GRB\,180325A is listed separately since it is not part of the F09 or XS-GRB sample papers. \\
			\tablefoottext{$\star$}{GRB afterglows where a detection of H$_2$ in absorption has been reported in the literature.} \\
			\tablefoottext{$\dagger$}{Bursts with a prominent detection of the 2175\,\AA~dust extinction feature.} \\
			\tablefoottext{$*$}{Tentative detection of C\,\textsc{i} below the imposed detection threshold.} \\
			\tablefoottext{$\ddagger$}{$A_V$ measurements from this work.}}
		\tablebib{The H\,\textsc{i} column densities, metallicities and visual extinctions are from: (1)~\citet{Delia07}; (2)~\citet{Zafar11a}; (3)~\citet{Ledoux09}; (4)~\citet{Schady12}; (5)~\citet{Covino13}; (6)~\citet{Cucchiara15}; (7)~\citet{Fynbo09}; (8)~\cite{Kann10}; (9)~\citet{Eliasdottir09}; (10)~\citet{DeCia12}; (11)~\citet{Zafar13}; (12)~\citet{Zafar12}; (13)~\citet{Prochaska09}; (14)~\citet{Delia10}; (15)~\citet{Zafar18a}; (16)~\citet{Wiseman17}; (17)~\citet{Delia14}; (18)~\citet{Kruehler13}; (19)~\citet{Friis15}; (20)~\citet{Zafar18c}; (21)~\citet{Hartoog15}; (22)~\citet{Selsing18}; (23)~Th\"one et al. (in preparation); (24)~Pugliese et al. (in preparation); (25)~\citet{DeUgartePostigo18}; (26)~\citet{Zafar18b}.}
		\label{tab:obs}
	\end{minipage}
\end{table*}

\subsection{Observations and sample compilation}\label{ssec:obs}

The majority of the GRB afterglows in our sample were detected by the Burst Alert Telescope (BAT) mounted on the Neil Gehrels \textit{Swift} Observatory \citep[\textit{Swift};][]{Gehrels04}. The few exceptions were detected by the \textit{Fermi} Gamma-Ray Space Telescope \citep{Atwood09,Meegan09} or by the \textit{INTEGRAL} satellite \citep{Winkler03}. 

The GRB afterglows from the F09 sample were observed using a range of low- to high-resolution spectrographs ($\mathcal{R} \approx 300 - 45\,000$), but we only include GRBs observed with $\mathcal{R} > 2000$ (see below). The more recent GRB afterglow sample by \cite{Selsing18} was obtained homogenously with the VLT/X-shooter spectrograph \citep{Vernet11}, covering a broad spectral range (300 -- 2480 nm) in a single exposure by splitting the light into three spectroscopic arms (called the UVB, VIS and NIR ams). We refer the reader to the sample papers for a description of the observational details and the specifics of the data reduction. All the spectra used in our analysis have been corrected for the foreground Galactic extinction using the dust maps of \cite{Schlegel98} and \cite{Schlafly11}. All wavelengths throughout the paper are reported in vacuum and are shifted to the heliocentric reference frame.

\subsection{Selection criteria and C\,\textsc{i} line identification} \label{ssec:sel}

We only include GRBs from the F09 and the XS-GRB sample for which there is spectral coverage of the C\,\textsc{i}\,$\lambda\lambda$\,1560,1656 line transitions. To exclude afterglow spectra that are host dominated and have poor signal-to-noise ratios (S/N) we imposed the following observational and brightness contraints for the GRB follow-up observations
\begin{enumerate}
	\item The GRB was observed within 24 hours after the trigger.
	\item The brightness measured from the acquisition image is brighter than $\leq 21.0$ mag.
	\item The rest-frame detection limit of C\,\textsc{i}\,$\lambda$\,1560 is 0.2\,\AA~(at 3$\sigma$). 
	\item The spectral resolution $\mathcal{R}$ is higher than 2000.
\end{enumerate}
The third criterion is defined such that we exlude bursts for which the upper limit on $W_{\mathrm{r}}(\lambda\,1560)$ cannot be constrained within 0.2\,\AA. We thus define this as our completeness limit and only consider GRB C\,\textsc{i} absorbers with $W_{\mathrm{r}}(\lambda\,1560) > 0.2$\,\AA~in our statistical sample.
The fourth criterion is only relevant for the F09 afterglow sample and is imposed to exclude the large fraction of bursts with $\mathcal{R} \approx 300$ for which C\,\textsc{i} is practically impossible to detect even at large S/N. We adopt the C\,\textsc{i} line detections and equivalent width measurements from the F09 paper, but measure the upper limits on C\,\textsc{i} for the rest of the GRBs entering our sample. Most bursts from F09 included in our sample already have derived upper limits on the column density of C\,\textsc{i} down to deep limits by \cite{Ledoux09}, but we report them in terms of equivalent widths here for consistency.

Since we require the spectral regions of the C\,\textsc{i}\,$\lambda\lambda$\,1560,1656 transitions to be covered, this effectively results in a redshift lower limit of $z_{\mathrm{min}} = \lambda_{\mathrm{min}}/1560 - 1 \approx 1.0 - 1.2$ for a spectral cut-off, $\lambda_{\mathrm{min}}$, of 300 to 350 nm. For typical optical spectrographs, where observations up to around 900 nm is possible, the redshift upper limit is $z_{\mathrm{max}}\approx5.8$, whereas for the VLT/X-shooter spectrograph, both C\,\textsc{i} line transitions can in principle be observed up to $z_{\mathrm{max}}\approx15$ (though with gaps due to the atmospheric transparency in the NIR). We are therefore able to probe a much larger redshift range than is possible for e.g. C\,\textsc{i} absorbers toward quasars observed as part of the SDSS survey \citep{Ledoux15}. To measure the chemical abundances of the GRB host galaxies (such as the neutral hydrogen column density, metallicity and dust depletion), however, requires that the GRBs are located at $z\gtrsim 1.7$ for the spectra to encompass the wavelength region where the Ly$\alpha$ absorption line can be observed from the ground.

%%%%%%%%%%%%%%%%%%%%%%%%%%%%%%%%%%%%%%%%%%%%%%%%%%%%%%%%%%%%%%%%%%%%%%%%%%%%
\section{Data analysis and sample properties}    \label{sec:met}
%%%%%%%%%%%%%%%%%%%%%%%%%%%%%%%%%%%%%%%%%%%%%%%%%%%%%%%%%%%%%%%%%%%%%%%%%%% 

In the final GRB afterglow sample, a total of 29 bursts (12/17 from the F09 and the XS-GRB samples, respectively) fulfill our imposed selection criteria and thus constitute our parent sample. Below we detail our measurements of the C\,\textsc{i}\,$\lambda\lambda$\,1560,1656 equivalent widths for the detections and tentative detections in the XS-GRB sample \citep[see][for the C\,\textsc{i} line detections and measurements from their sample]{Fynbo09}. We also include GRB\,060210 from the F09 sample as a C\,\textsc{i} absorber in our sample even though C\,\textsc{i}\,$\lambda$\,1656 is not detected, due to its wavelength region being outside the spectral coverage, since the other transitions C\,\textsc{i}\,$\lambda\lambda\lambda$\,1260,1277,1328 are detected. We report the full set of C\,\textsc{i}\,$\lambda\lambda$\,1560,1656 equivalent width measurements and the derived $3\sigma$ upper limits for the GRBs with non-detections of C\,\textsc{i} in Table~\ref{tab:obs}.

\begin{figure} [!t]
	\centering
	\epsfig{file=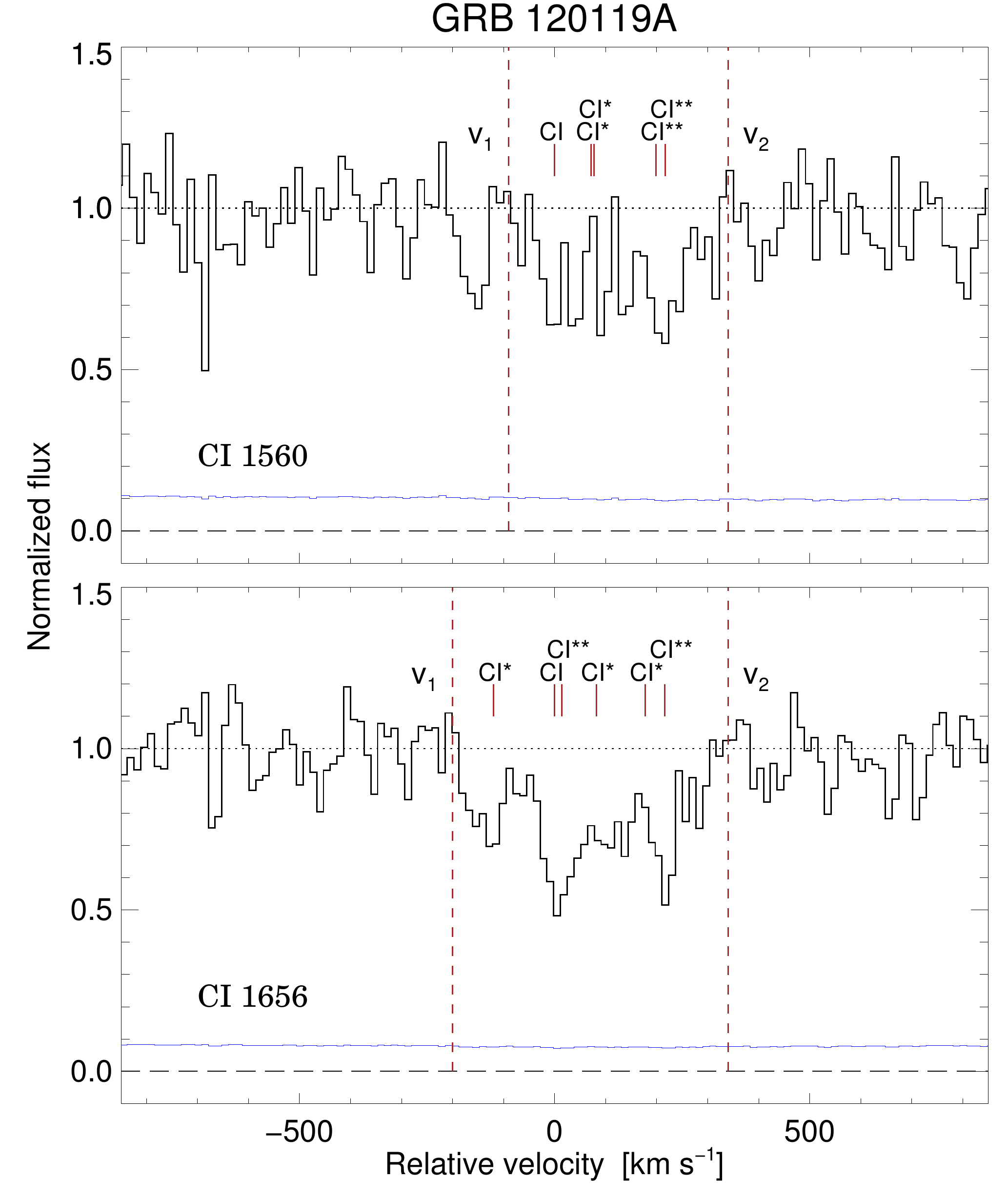,width=\columnwidth}
	\caption{VLT/X-shooter spectrum of the GRB\,120119A in velocity space, centred on the ground-state of the C\,\textsc{i}\,$\lambda\lambda$\,1560,1656 absorption lines at $z=1.7291$. The black solid line shows the spectrum and the associated error is shown in blue. The spectrum has been binned by a factor of two. The ground-state and fine-structure lines of the C\,\textsc{i}\,$\lambda\lambda$\,1560,1656 transitions are marked above each of the absorption profiles. The profiles have been integrated over the velocity range indicated by the dashed red lines from $v_1$ to $v_2$ to measure the equivalent widths. C\,\textsc{i} is detected above the completeness limit in this system.}
	\label{fig:120119a_cispec}
\end{figure}

\subsection{Equivalent width of C\,\textsc{i}} \label{ssec:ew}

For all GRBs we provide the total rest-frame equivalent widths of the C\,\textsc{i}\,$\lambda\lambda$\,1560,1656 absorption lines and list them in Table~\ref{tab:obs}. We attribute all the absorption to the ground-state of C\,\textsc{i}, even though the fine-structure transitions C\,\textsc{i}* and C\,\textsc{i}** also contribute \citep{Srianand05,Noterdaeme17}. This is, however, to be consistent with the measurements from the F09 sample and to directly compare our results to the sample of quasar C\,\textsc{i} absorbers from \cite{Ledoux15}. We determine the rest-frame equivalent widths of the C\,\textsc{i}\,$\lambda\lambda$\,1560,1656 absorption features by fitting the continuum around each of the lines and then integrate the absorption profile contained below the normalized flux level which encompasses all the C\,\textsc{i}\,$\lambda\lambda$\,1560,1656 ground-state and fine-structure transitions.

\subsubsection{GRB\,120119A}

The spectrum of GRB\,120119A at $z=1.7288$ is presented in the work by \cite{Wiseman17}. We adopt the H\,\textsc{i} column density of $\log N$(H\,\textsc{i}/cm$^{-2}$) = $22.44\pm 0.12$ and the derived metallicity [Zn/H] = $-0.96\pm 0.28$ from their work. \cite{Zafar18a} measured a visual extinction of $A_V = 1.02\pm 0.11$ mag and found no evidence for the presence of the 2175\,\AA~dust extinction feature. In Fig.~\ref{fig:120119a_cispec}, we show a part of the spectrum centred on the C\,\textsc{i}\,$\lambda\lambda$\,1560,1656 line transitions. The absorption profiles have been integrated over the velocity range indicated by the dashed red lines from $v_1$ to $v_2$ to measure the equivalent widths and include the ground-state and fine-structure lines. We measure total C\,\textsc{i} rest-frame equivalent widths of $W_\mathrm{r}(\lambda\,1560) = 0.51\pm 0.05$\,\AA~and $W_\mathrm{r}(\lambda\,1656) = 0.77\pm 0.06$\,\AA.

\begin{figure} [!t]
	\centering
	\epsfig{file=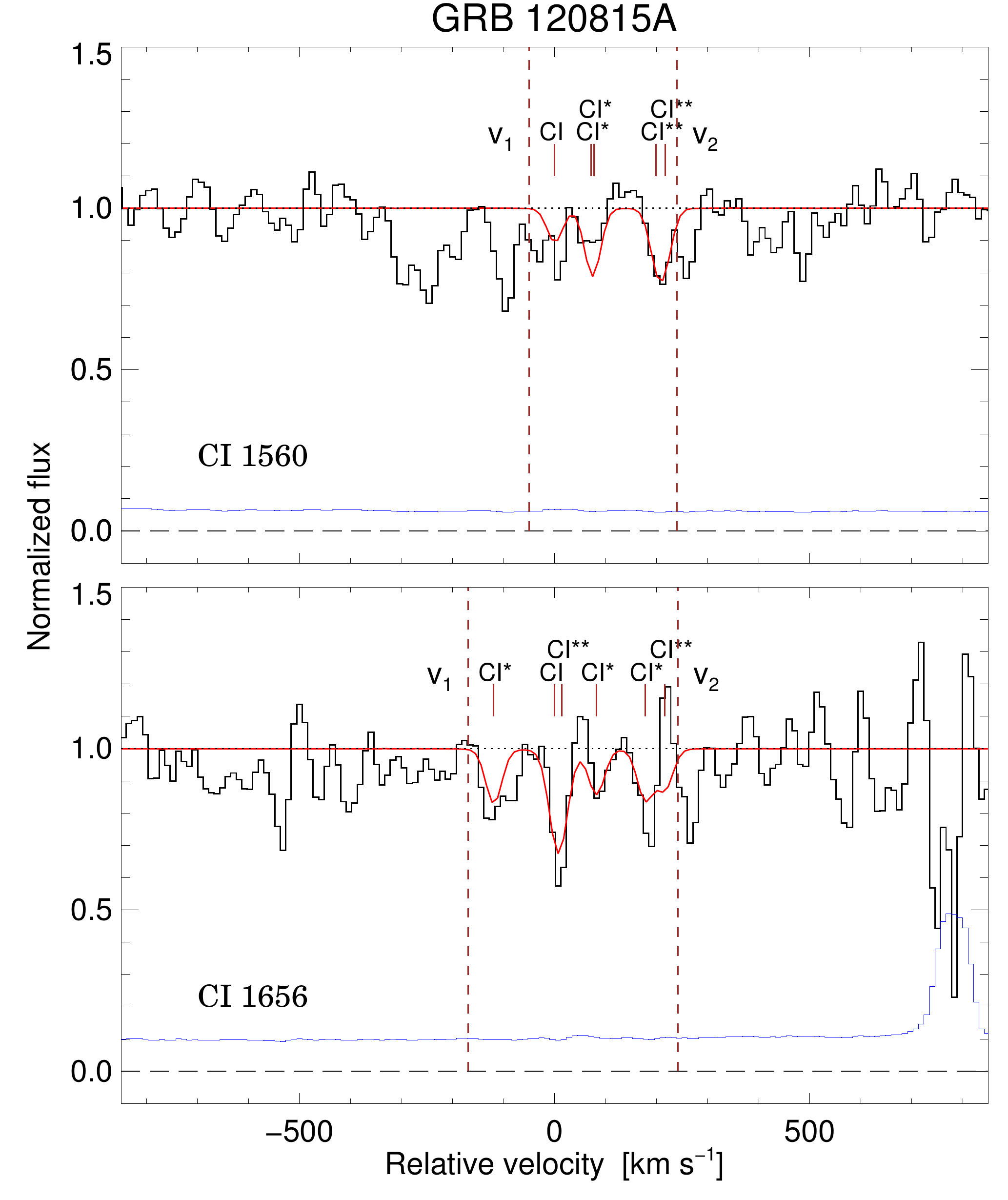,width=\columnwidth}
	\caption{Same as Fig.~\ref{fig:120119a_cispec} but for GRB\,120815A, centred on $z=2.3581$. Consistent absorption profiles are also detected in C\,\textsc{i}\,$\lambda$\,1277 and C\,\textsc{i}\,$\lambda$\,1328. Overplotted in red are the best-fit Voigt profiles. C\,\textsc{i} is detected below the completeness limit in this system.}
	\label{fig:120815a_cispec}
\end{figure}

\subsubsection{GRB\,120815A}

\begin{figure} [!t]
	\centering
	\epsfig{file=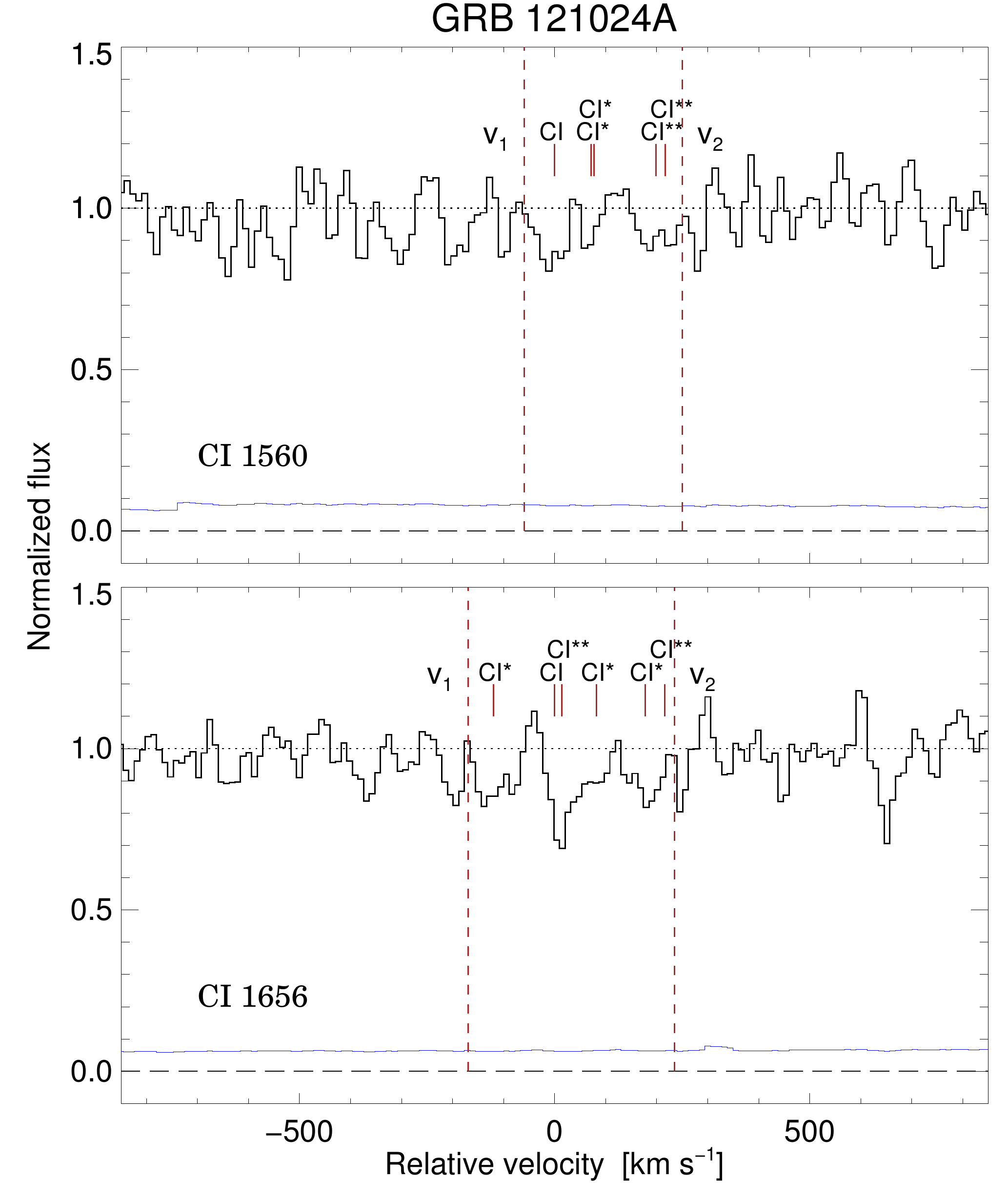,width=\columnwidth,height=10.4cm}
	\caption{Same as Fig.~\ref{fig:120119a_cispec} but for GRB\,121024A, centred on $z=2.3021$. C\,\textsc{i} is detected below the completeness limit in this system.}
	\label{fig:121024a_cispec}
\end{figure}

\begin{figure} [!t]
	\centering
	\epsfig{file=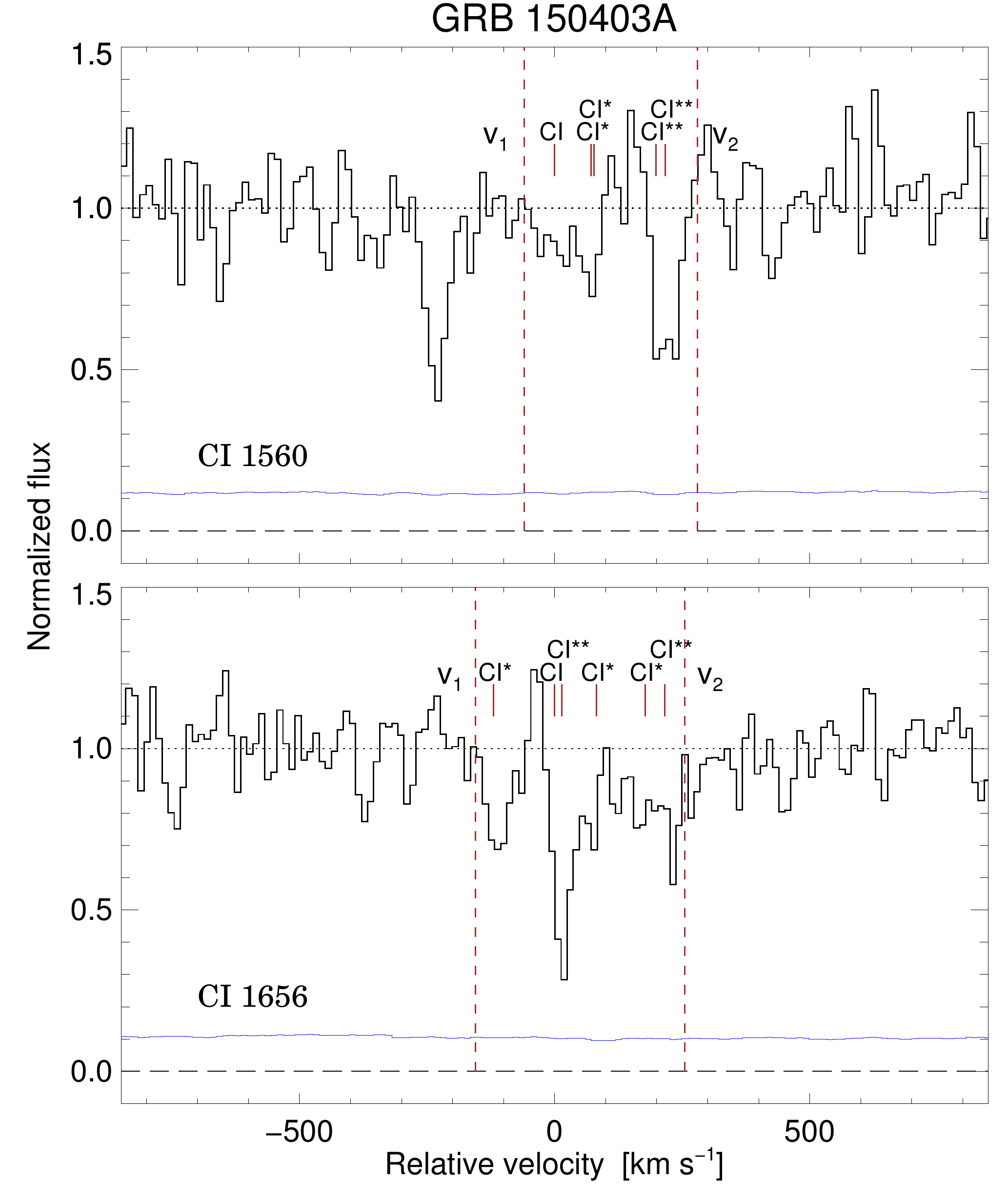,width=\columnwidth,height=10.4cm}
	\caption{Same as Fig.~\ref{fig:120119a_cispec} but for GRB\,150403A, where the ground-state transitions are centred on $z=2.0569$. C\,\textsc{i} is detected above the completeness limit in this system.}
	\label{fig:150403a_cispec}
\end{figure}

The spectrum of GRB\,120815A at $z=2.3581$ is presented in the work by \cite{Kruehler13}. We adopt the H\,\textsc{i} column density of $\log N$(H\,\textsc{i}/cm$^2$) = $21.95\pm 0.10$ and the derived metallicity [Zn/H] = $-1.15\pm 0.12$ from their work. They also detect absorption lines from H$_2$ with a column density of $\log N$(H$_2$/cm$^2$) = $20.54\pm 0.13$, which yields a molecular gas fraction of $\log f$(H$_2$) = $-1.14\pm 0.15$, and C\,\textsc{i} with a column density of $\log N$(C\,\textsc{i}/cm$^2$) = $13.41\pm 0.11$. \cite{Zafar18a} measured a visual extinction of $A_V = 0.19\pm 0.04$ mag and found no evidence for the presence of the 2175\,\AA~dust extinction feature. We measure equivalent widths of $W_\mathrm{r}(\lambda\,1560) = 0.12\pm 0.08$\,\AA~and $W_\mathrm{r}(\lambda\,1656) = 0.21\pm 0.05$\,\AA~(shown in Fig.~\ref{fig:120815a_cispec}). The equivalent width of C\,\textsc{i}\,$\lambda$\,1560 is below the detection threshold of $W_{\mathrm{r}}^{\mathrm{lim}} = 0.2\,\AA$, however, and we therefore only report the C\,\textsc{i} in this system as a tentative detection. To verify the detection we compared the velocity components seen for the C\,\textsc{i}\,$\lambda\lambda$\,1560,1656 lines to the normalized flux at the region of the C\,\textsc{i}\,$\lambda\lambda$\,1277,1328 line transitions and found consistent results. Based on these four sets of C\,\textsc{i} ground-state and fine-structure lines we fit Voigt profiles to all components and line transitions to further demonstrate the robustness of the detection. The best fit is shown in Fig.~\ref{fig:120815a_cispec} as the red solid line. We then also measured the equivalent widths directly from the Voigt-profile model and found consistent results. 

\subsubsection{GRB\,121024A}

The spectrum of GRB\,121024A at $z=2.3024$ is presented in the work by \cite{Friis15}. We adopt the H\,\textsc{i} column density of $\log N$(H\,\textsc{i}/cm$^2$) = $21.88\pm 0.10$ and the derived metallicity [Zn/H] = $-0.70\pm 0.10$ from their work. They also detect absorption lines from H$_2$ with a column density of $\log N$(H$_2$/cm$^2$) $\approx 19.8$, which yields a molecular gas fraction of $\log f$(H$_2$) $\approx -1.4$. They do not report a detection of C\,\textsc{i}, but after re-examing the spectrum we identify the ground-state and fine-structure lines belonging to the C\,\textsc{i}\,$\lambda\lambda$\,1560,1656 transitions (see Fig.~\ref{fig:121024a_cispec}). \cite{Zafar18a} measured a visual extinction of $A_V = 0.19\pm 0.04$ mag for this GRB and found no evidence for the presence of the 2175\,\AA~dust extinction feature. We measure equivalent widths of $W_\mathrm{r}(\lambda\,1560) = 0.08\pm 0.05$\,\AA~and $W_\mathrm{r}(\lambda\,1656) = 0.11\pm 0.07$\,\AA. The equivalent width of C\,\textsc{i}\,$\lambda$\,1560 is below the detection threshold of $W_{\mathrm{r}}^{\mathrm{lim}} = 0.2\,\AA$, however, and we therefore only report the C\,\textsc{i} in this system as a tentative detection.

\begin{figure} [!t]
	\centering
	\epsfig{file=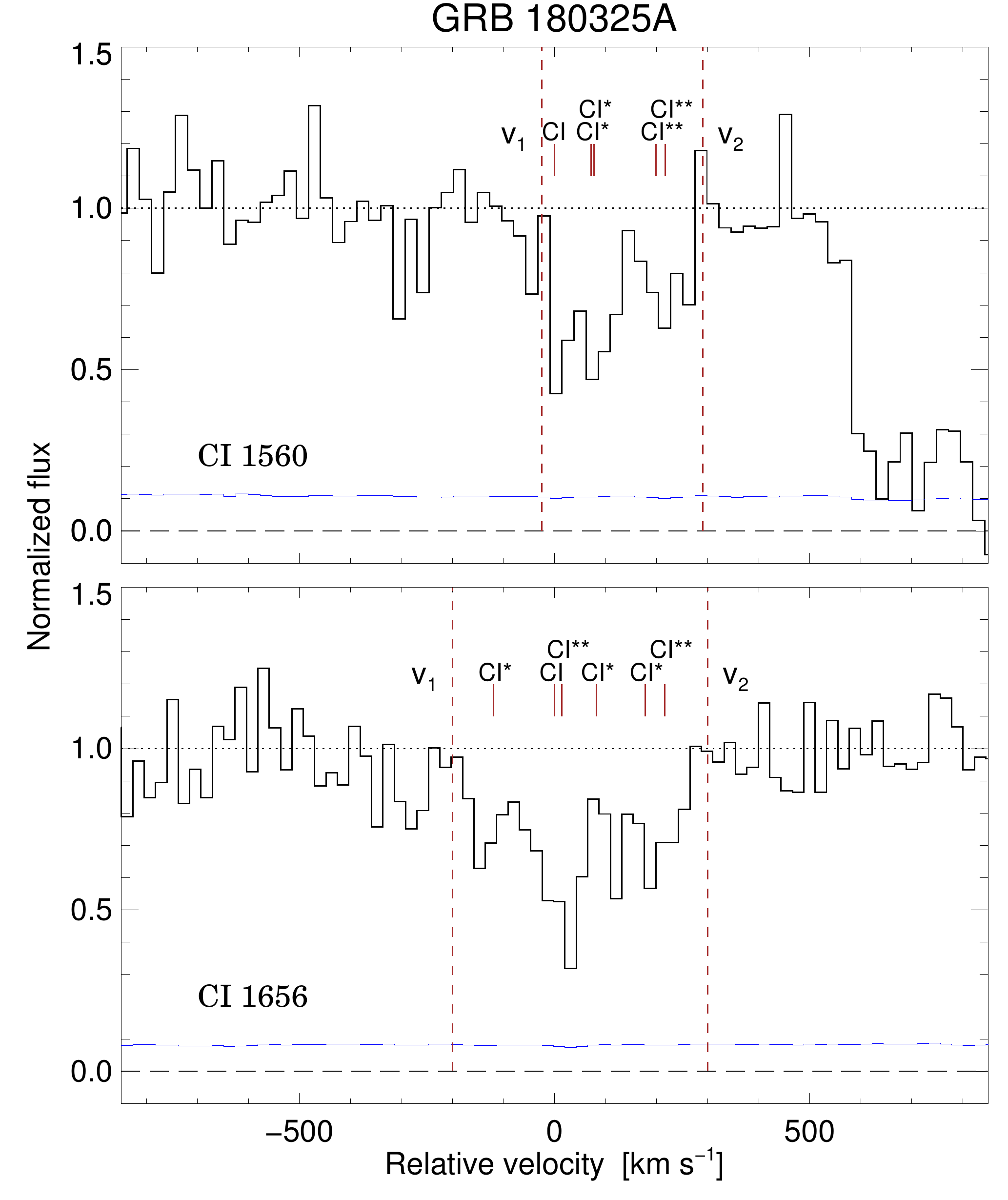,width=\columnwidth}
	\caption{Same as Fig.~\ref{fig:120119a_cispec} but for GRB\,180325A, where the ground-state transitions are centred on $z=2.2494$. C\,\textsc{i} is detected above the completeness limit in this system. The spectrum has been binned by a factor of two.}
	\label{fig:180325a_cispec}
\end{figure}

\subsubsection{GRB\,150403A}

The spectrum of GRB\,150403A at $z=2.0571$ is presented in the work by \cite{Selsing18} from which we adopt the H\,\textsc{i} column density of $\log N$(H\,\textsc{i}/cm$^2$) = $21.80\pm 0.20$. Detailed measurements of the gas-phase abundances for this burst will be presented in the work by Th\"one et al. (in preparation), but see also \cite{Heintz18a} where they report [S/H] = $-0.80\pm 0.35$. In this work we measure a visual extinction of $A_V = 0.20\pm 0.03$ mag (see below), and find no evidence for the presence of the 2175\,\AA~dust extinction feature. The C\,\textsc{i}\,$\lambda\lambda$\,1560,1656 line transitions are readily detected in the spectrum (see Fig.~\ref{fig:150403a_cispec}) and we measure equivalent widths of $W_\mathrm{r}(\lambda\,1560) = 0.34\pm 0.03$\,\AA~and $W_\mathrm{r}(\lambda\,1656) = 0.50\pm 0.04$\,\AA.

\subsubsection{GRB\,180325A}

The spectrum of GRB\,180325A at $z=2.2486$ is presented in the work by \cite{Zafar18b}. We adopt the H\,\textsc{i} column density of $\log N$(H\,\textsc{i}/cm$^2$) = $22.30\pm 0.14$, the derived metallicity [Zn/H] = $>-0.98$, the measured visual extinction, $A_V = 1.58\pm 0.12$ mag, and the measured C\,\textsc{i} equivalent widths $W_\mathrm{r}(\lambda\,1560) = 0.58\pm 0.05$\,\AA~and $W_\mathrm{r}(\lambda\,1656) = 0.85\pm 0.05$\,\AA~from their work (see also Fig.~\ref{fig:180325a_cispec}). They also detect a strong 2175\,\AA~dust extinction feature, the strength of which follow the correlation expected from the C\,\textsc{i} equivalent widths \citep{Ma18}.

\subsection{Dust extinction}  \label{ssec:av}

Due to their immense luminosity and well-known intrinsic power-law spectra, GRBs are an ideal probe of dust in star forming galaxies and in the line of sight out to high redshifts \citep{Zafar11a,Zafar11b,Greiner11,Covino13,Bolmer18a}. GRB extinction curves can typically be described by similar prescriptions as that of the Small Magellanic Cloud (SMC), but can in rare cases also show a prominent 2175\,\AA~extinction bump known locally from sightlines in the Milky Way (MW) and towards the Large Magellanic Cloud (LMC) \citep[e.g.][]{Kruehler08,Eliasdottir09,Prochaska09,Perley11,Zafar12,Zafar18b}. GRB afterglows have recently been found to show slightly steeper extinction curves on average than that of e.g. the SMC \citep{Zafar18a}. Even more unusual extinction features have also been observed, such as flat or "grey" \citep{Savaglio04,Perley08,Friis15} or very steep \citep{Fynbo14,Heintz17} reddening curves. By utilizing the simple and smooth intrinsic power-law spectra of GRBs, the specific extinction curves can be well-constrained and provide a measure of the visual extinction, $A_V$, in the line of sight to the GRB, in addition to the overall dust composition and grain size distribution.

The majority of the GRB afterglows in our parent sample already have published measurements of the extinction. Here we provide values of $A_V$ for four additional bursts (GRBs\,141109A, 150403A, 151021A, and 160203A) observed with VLT/X-shooter (see Table~\ref{tab:obs}) using the approach detailed in the sections below. Our procedure is similar to previous determinations of $A_V$'s in GRBs from the literature, although we do not rely on multi-epoch photometry to normalize the X-ray to the optical spectra. We find consistent results using our approach, compared to already published values of $A_V$ in GRB sightlines.

\subsubsection{Intrinsic afterglow SED}

The continuum emission from a GRB afterglow is believed to be dominated by synchrotron radiation described by a single or broken power-law \citep{Sari98}. It has also been verified observationally that the intrinsic X-ray spectrum derived from the \textit{Swift}/XRT can be used as proxy for the intrinsic optical spectrum, typically with a cooling break of $\Delta\beta=0.5$ \citep{Zafar11a}. The X-ray spectral slope in photon units, $\Gamma$, is given by a fit to the \textit{Swift}/XRT spectrum in the dedicated repository\footnote{\url{http://www.swift.ac.uk/xrt_spectra/}}. From the photon index, the prescription for the intrinsic optical spectral slope as a function of wavelength is given as
\begin{equation}
F_{\lambda} = F_0\,\lambda^{(\Gamma - \Delta\beta - 3)}~,
\label{eq:int}
\end{equation}
or as a function of frequency, $F_{\nu} = F_0\,\nu^{(\Gamma - \Delta\beta - 2)}$.
In the following analysis we will use the photon index from the \textit{Swift}/XRT database to derive the intrinsic spectral slope and allow for the slope change due to the cooling break to take a value to take a value of $\Delta\beta=0.0$ or 0.5. 

Ideally, the optical/NIR and X-ray spectral energy distributions (SEDs) should be fit together to get an estimate of the intrinsic slope \citep[see e.g.][]{Watson06,Greiner11,Schady12,Japelj15,Bolmer18a}. To do so, however, requires that multiple photometric data points are available to normalize the optical and X-ray spectra to the same epoch using their respective lightcurves. A subset of the GRBs observed with VLT/X-shooter that have coinciding multi-epoch photometry have already been published by \cite{Japelj15} and \cite{Zafar18a}. Using Eq.~\ref{eq:int} we can estimate the $A_V$ in an alternative way, not relying on near-simultaneous photometric observations. To improve the absolute flux calibration of the VLT/X-shooter spectra, although not particularly important for our approach, we rescale the VIS arm to the measured acquisition magnitude (typically $R$-band). The flux of the GRB in the UVB and NIR arms are then scaled to match the VIS arm spectrum.

\subsubsection{Dust-extinction model}

The observed optical/NIR flux from the GRB afterglow is extinguished due to absorption or scattering by dust particles located in the line of sight to the burst. In the majority of cases the dominant contribution is from dust in the GRB host galaxy (after correcting for the Galactic extinction). The observed afterglow spectrum can therefore be described as
\begin{equation}
F_{\lambda}^{\mathrm{obs}} = F_{\lambda}\times 10^{-0.4\,A_{\lambda}}~,
\end{equation}
where $A_{\lambda}$ is the extinction as a function of wavelength, $\lambda$, in the GRB host galaxy. The visual extinction, $A_V$, can then be estimated assuming a given dust-extinction model when knowing the redshift of the GRB and using the intrinsic spectral shape as measured from the X-ray spectrum.

We fit the combined X-shooter spectrum with the dust-extinction model of \cite{Fitzpatrick90} which parametrizes the extinction curve through a set of eight parameters. The extinction curve is defined as
\begin{equation}
A_{\lambda} = \frac{A_V}{R_V}\left(k(\lambda - V) + 1\right)~,
\end{equation}
where the relative reddening, $k(\lambda - V)$, is given as
\begin{equation}
k(\lambda - V) = c_1+c_2x+c_3D(x,x_0,\gamma) + c_4F(x)~,
\end{equation}
with 
\begin{equation}
F(x) = \begin{cases} 
0.539(x-5.9)^2 + 0.056(x-5.9)^3 &\mathrm{for}~x\ge 5.9  \\
0 &\mathrm{for}~x< 5.9
\end{cases}~,
\end{equation}
and the Lorentzian-like Drude profile representing the 2175\,\AA~extinction bump, observed locally in some sightlines in the MW and the LMC, is described as
\begin{equation}
D(x,x_0,\gamma) = \frac{x^2}{(x^2-x_0^2)^2+x^2\gamma^2}~,
\end{equation}
where $x = (1~\mu\mathrm{m})/\lambda$. Basically, this dust-extinction model contains two components, one describing the linear UV part of the spectrum via the components $c_1$ (intercept), $c_2$ (slope) and the term $c_4F(x)$ describing the far-UV curvature. The second component is the Drude profile, controlled by the parameters $c_3$ (bump strength), $x_0$ (central wavelength) and $\gamma$ (width of the bump). The last two parameters are the visual extinction, $A_V$, and the total-to-selective reddening, $R_V$. The advantage of such a parametrization of the dust-extinction model is that it allows for, e.g., the strength and width of the 2175\,\AA~bump to be fitted independently \citep{Jiang10,Ledoux15,Ma17,Noterdaeme17} and to model extinction curves with no local analogs \citep{Fynbo14,Amanullah14,Heintz17}. We do not observe any evidence for the 2175\,\AA~extinction bump or an unusual steep (or flat) reddening curve in the four GRBs examined in this work.

\begin{figure} [!t]
	\centering
	\epsfig{file=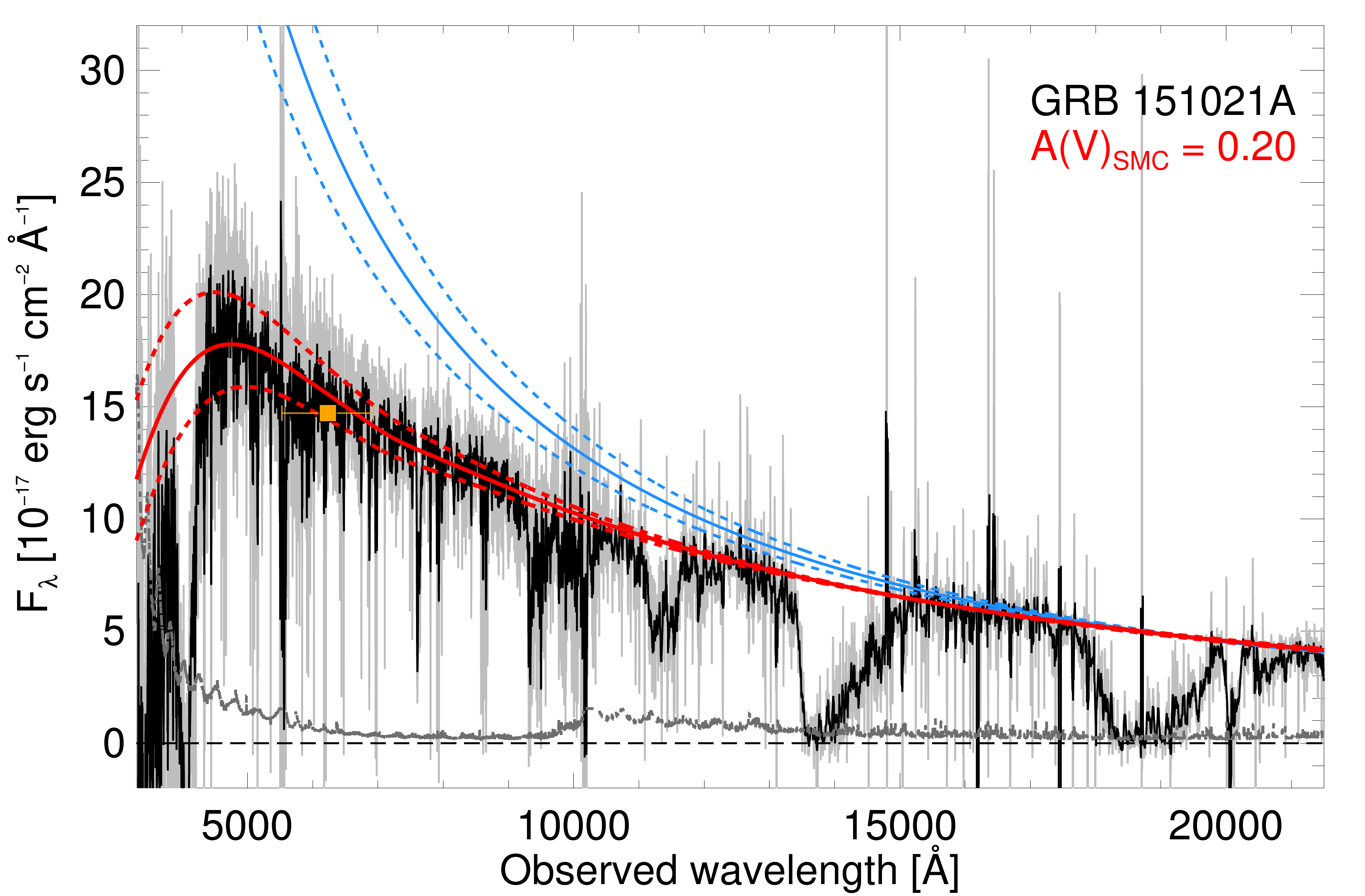,width=\columnwidth}
	\caption{Fit of the dust extinction, $A_V$, to the spectrum of GRB\,151021A. The grey and black lines show the full raw and binned X-shooter spectrum, respectively. The corresponding raw error spectrum is shown as the bottom dark grey line. The spectrum was normalized to the acquisition magnitude overplotted as the orange square. The intrinsic afterglow spectrum derived from the \textit{Swift}/XRT spectrum with a slope change of $\Delta\beta =0.5$ is shown as the blue solid line with the errors on the power-law slope shown as the blue, dashed lines. An extinction curve with the best fit value of $A_V=0.20\pm 0.03$ mag is shown as the red, solid line and the error on the fit is shown by the red, dashed lines.}
	\label{fig:extfit}
\end{figure}

We derive the rest-frame visual extinction, $A_V$, by normalizing the intrinsic power-law spectrum to the flux in the NIR arm in the wavelength region of a typical $K$-band. We then fit the observed spectrum using three different reddening laws: SMC and LMC as parametrized by \cite{Gordon03} and the slightly steeper reddening law inferred for the average GRB afterglow derived by \cite{Zafar18a}, all using the \cite{Fitzpatrick90} prescription. We fix the redshift of the dust component to $z_{\mathrm{GRB}}$ and then only vary $A_V$. In Fig.~\ref{fig:extfit} we show an example of one of the fits for GRB\,151021A. In this case we measure $A_V = 0.20\pm 0.03$ mag and find that the extinction is best fit with the SMC reddening law. To verify our approach we also fit the XS-GRBs for which the visual extinction has already been measured and find consistent results. Our derived $A_V$ values are provided in Table~\ref{tab:obs}.

\begin{figure} [!t]
	\centering
	\epsfig{file=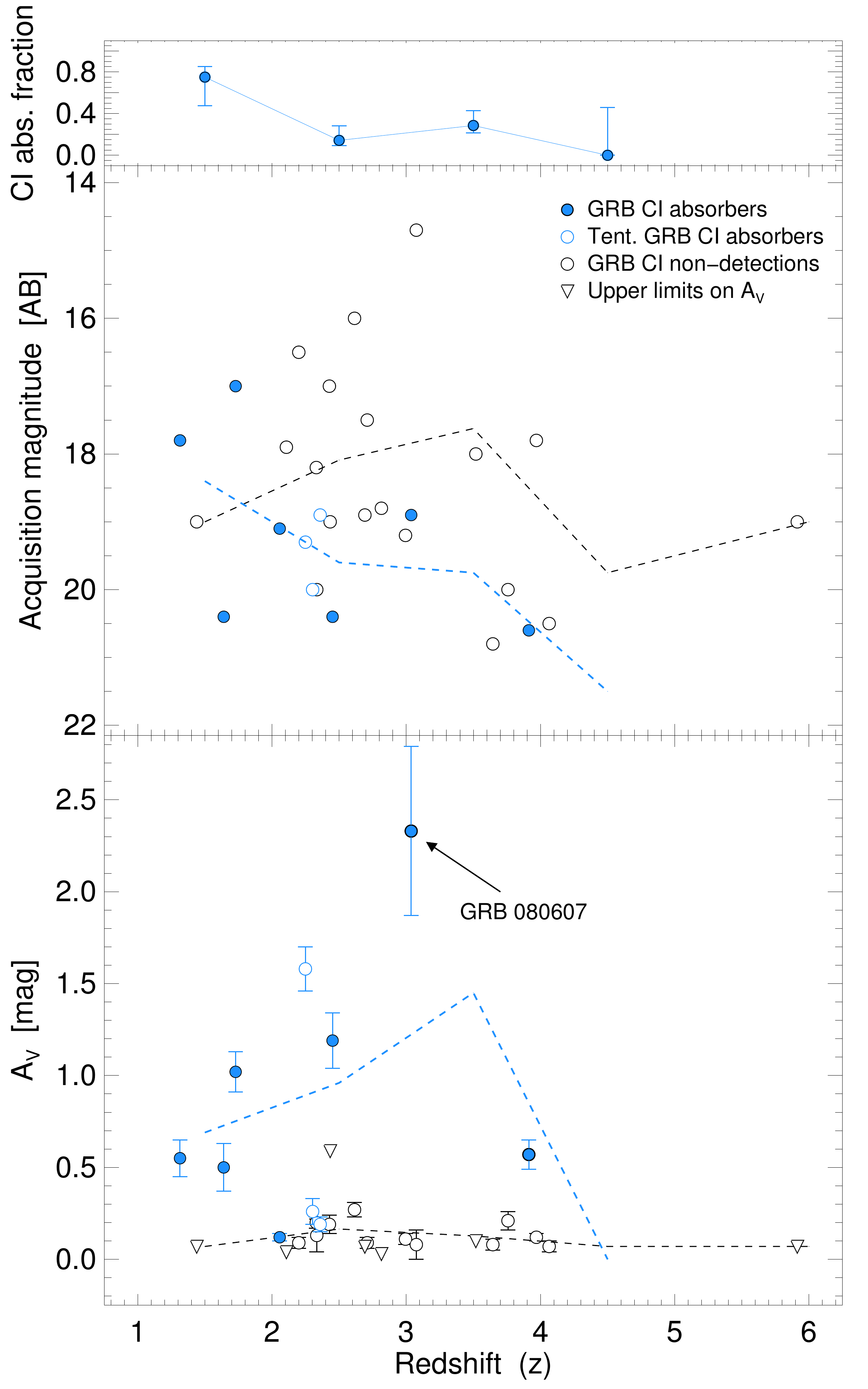,width=1.0\columnwidth}
	\caption{Redshift distribution as a function of GRB brightness at the time of observation (middle panel) and dust extinction, $A_V$ (bottom panel). Blue filled circles denote the GRBs in our statistical sample with C\,\textsc{i} detected in absorption. Black empty circles mark the GRBs in our sample without C\,\textsc{i}, where triangles represent upper limits on $A_V$. In the middle and bottom panels, the dashed lines show the mean acquisition magnitudes and $A_V$ values for the GRBs without (black) and with (blue) C\,\textsc{i} detected in absorption in different bins: $1 < z < 2$, $2 < z < 3$, $3 < z < 4$, and $z > 4$. In the top panel the fraction of GRB C\,\textsc{i} absorbers is shown in the same redshift interval. The GRBs with C\,\textsc{i} detected below the completeness limit are shown by the blue empty circles, and are not included in the analysis.}
	\label{fig:ciz}
\end{figure}

\subsection{Sample properties}

The C\,\textsc{i} detections are reported in Table~\ref{tab:obs}. It is detected robustly (i.e. with $W_{\mathrm{r}}(\lambda\,1560) > 0.2$\,\AA) in eight GRBs, five from the F09 sample (GRBs\,060210, 060607A, 061121, 080605, and 080607) and three from the XS-GRB sample (GRBs\, 120119A, 150403A, and 180325A). We do not include GRB\,180325A in the following statistical analysis since it is not part of the two sample papers and was added specifically because of its detection of C\,\textsc{i}. In the statistical sample, seven of
the GRB afterglows ($\approx 25\%$) thus have C\,\textsc{i} detected in absorption, with an approximately $40\%$ detection rate in the F09 sample and $\approx15\%$ in the XS-GRB sample. The reason for the larger detection in the F09 sample is likely due to most of the afterglows have been obtained with more sensitive, low-resolution spectrographs compared to the bursts observed with the VLT/X-shooter instrument. Due to the usage of more sensitive, low-resolution spectrographs pre-X-shooter, a larger fraction of faint bursts that are potentially more obscured due to larger dust columns entered the F09 sample. As a consequence, this sample will contain more dust obscured
afterglows, and therefore a higher detection probability of C\,\textsc{i} is expected. This is also evident from Table~\ref{tab:obs}, showing that the bursts from the F09 sample have on average higher values of $A_V$ (with a mean of $A_V = 0.54$\,mag) than those from the XS-GRB sample (mean of $A_V = 0.18$\,mag). Furthermore, the two fractions are even consistent within $\approx 1\sigma$ when taking into account the low number statistics \citep{Cameron11}. We also identify absorption from C\,\textsc{i} in GRBs\,120815A and 121024A but below the sample completeness limit and we therefore do not include them in the statistical sample. 

Our full sample spans a redshift range of $z = 1 - 6$ with C\,\textsc{i} absorption detected up to $z\approx 4$ in GRB\,060210. C\,\textsc{i} is also detected in all GRBs where H$_2$ has previously been reported in absorption via the Lyman-Werner bands, except for GRB\,120327A (which has a low molecular fraction) for which we place upper limits of $W_{\mathrm{r}}(\lambda\,1560) < 0.03$\,\AA~and $W_{\mathrm{r}}(\lambda\,1656) < 0.03$\,\AA. This GRB is observed to have an insignificant amount of dust \citep[$A_V < 0.03$ mag,][]{Delia14}, however, which would be consistent with the non-detection of C\,\textsc{i} (see Sect.~\ref{sec:results}).

In general, there is a strong excess of cold gas in GRB hosts compared to absorbers in quasar sightlines \citep{Ledoux15}, where about one percent of intervening C\,\textsc{i} absorbers were found with a completeness of around 40\% at $W_{\mathrm{r}}^{\mathrm{lim}} = 0.2\,\AA$. The strong excess of cold gas is not unexpected since GRBs probe more central regions of their host galaxies whereas quasars probe random sightlines through intervening absorbers. Due to higher ISM pressure, the covering fraction of the cold, neutral gas is also expected to be higher closer to the galactic centre.

In the top panel of Fig.~\ref{fig:ciz} we show the fraction of GRB C\,\textsc{i} absorbers as a function of redshift, binned by $\Delta z = 1$ from $z = 1 - 4$ \citep[where the errors are calculated assuming small number statistics, see][]{Cameron11}. The GRB C\,\textsc{i} absorber fraction is high at $1 < z < 2$ with $\approx 75\%$, where for $z > 2$ the fraction ranges from $\approx 20\% - 30\%$. Here we only consider the C\,\textsc{i} detections from the statistical sample, thereby excluding GRBs\,120815A, 121024A, and 180325A (but include them in the figure as empty blue circles). In the middle panel of Fig.~\ref{fig:ciz}, the redshift distribution is shown as a function of the observed acquisition magnitude (typically in the $R$-band). We observe that the GRB afterglows with C\,\textsc{i} detected in absorption on average have fainter magnitudes. In the bottom panel of Fig.~\ref{fig:ciz} the redshift distribution is shown as a function of the visual extinction, $A_V$, for each burst. In general, the GRB C\,\textsc{i} absorbers are found to be the most dust-reddened systems at all redshifts, shown by the curves representing the mean of the two populations.

 \begin{figure*} [!t]
 	\centering
 	\epsfig{file=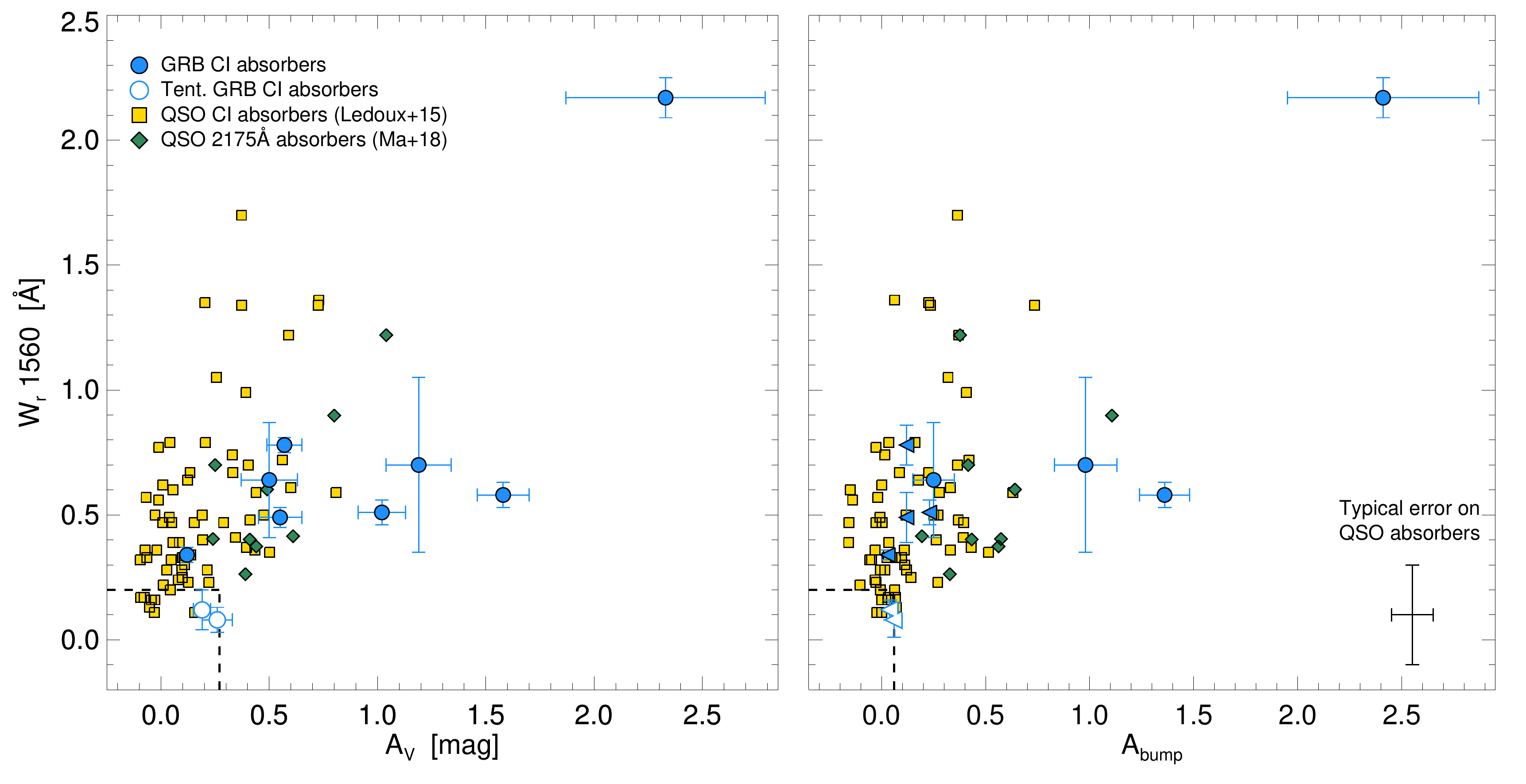,width=16cm,height=7.9cm}
 	\caption{C\,\textsc{i}\,$\lambda$\,1560 rest-frame equivalent width as a function of dust extinction, $A_V$ (left panel) and 2175\,\AA\ bump strength, $A_{\mathrm{bump}}$ (right panel). The blue filled circles denote the GRBs with C\,\textsc{i} detected in absorption (empty blue circles show the GRBs with C\,\textsc{i} detected below the completeness limit), where the filled blue left-pointing triangles in the right panel represent the upper limit on $A_{\mathrm{bump}}$ derived for the other GRB C\,\textsc{i} absorbers (assuming an SMC bar extinction curve). The yellow squares show the quasar C\,\textsc{i} absorbers from \cite{Ledoux15} and the green diamond symbols represent the quasar 2175\,\AA~dust extinction feature absorbers from \cite{Ma18}. The dashed lines mark the $W_\mathrm{r}(\lambda\,1560)$ detection limit and maximum $A_V$ and $A_{\mathrm{bump}}$ for the GRBs with non-detections of C\,\textsc{i}.}
 	\label{fig:w1560av}
 \end{figure*}

In our full sample of GRB C\,\textsc{i} absorbers, only four afterglows (GRBs\,070802, 080605, 080607, and 180325A) have a robust detection of the 2175\,\AA~dust extinction feature. All four have considerable visual extinctions \citep[$A_V > 0.5$ mag,][]{Zafar12,Zafar18b} and strong C\,\textsc{i} equivalent widths of $W_{\mathrm{r}}(\lambda\,1560) \gtrsim 0.6$\,\AA. Only three other GRBs with C\,\textsc{i} absorption in our full sample, namely GRBs\,060210, 061121 and 120119A, have $A_V > 0.5$ mag as well, but do not show any indication of the 2175\,\AA~dust extinction feature. For the GRBs\,060210 and 061121, \cite{Covino13} found a best fit with an SMC-like extinction curve, though only based on its broad-band photometry. Initially, \cite{Japelj15} found a best fit with an LMC-like extinction curve for GRB\,120119A but \cite{Zafar18a} argue that it was a false detection based on the unbinned X-shooter spectrum and found a best fit with an SMC-like extinction curve instead. In the statistical sample of GRB C\,\textsc{i} absorbers, about $40\%$ thus show the presence of the 2175\,\AA~dust extinction feature \citep[consistent with the quasar C\,\textsc{i} absorber sample;][]{Ledoux15}.

 \begin{figure} [!t]
 	\centering
 	\epsfig{file=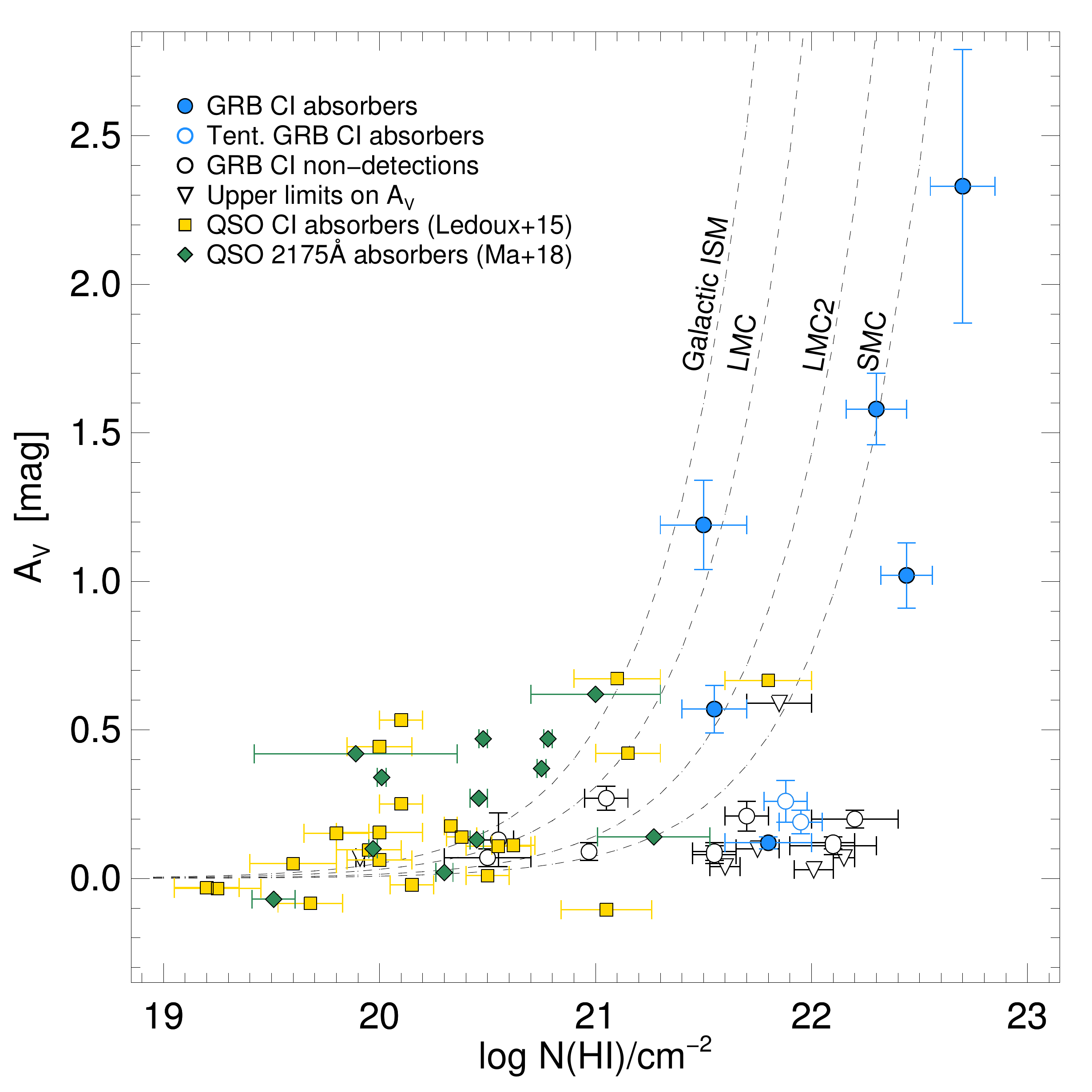,width=\columnwidth}
 	\caption{Column density of neutral hydrogen (H\,\textsc{i}) as a function of dust extinction, $A_V$, i.e. the dust-to-gas ratio. Blue filled circles again denote the GRBs with C\,\textsc{i} detected in absorption (empty blue circles show the GRBs with C\,\textsc{i} detected below the completeness limit). Black empty circles mark the GRBs in our samples without C\,\textsc{i}, where triangles represent bursts with upper limits on $A_V$. The yellow squares show the quasar C\,\textsc{i} absorbers from \cite{Ledoux15} and the green diamond symbols represent the quasar 2175\,\AA~dust extinction feature absorbers from \cite{Ma18}. Overplotted are the average dust-to-gas ratios from specific sightlines in the Local Group (MW, LMC, LMC2 and SMC) from \cite{Gordon03}.}
 	\label{fig:hiav}
 \end{figure}

%%%%%%%%%%%%%%%%%%%%%%%%%%%%%%%%%%%%%%%%%%%%%%%%%%%%%%%%%%%%%%%%%%%%%%%%%%%%
\section{Results}    \label{sec:results}
%%%%%%%%%%%%%%%%%%%%%%%%%%%%%%%%%%%%%%%%%%%%%%%%%%%%%%%%%%%%%%%%%%%%%%%%%%%

The left panel of Fig.~\ref{fig:w1560av} shows the C\,\textsc{i} equivalent width as a function of the visual extinction, $A_V$, for the GRB C\,\textsc{i} absorbers. We also overplot the quasar C\,\textsc{i} absorbers from \cite{Ledoux15} and the 2175\,\AA~dust extinction feature absorbers, all with C\,\textsc{i} detected in absorption as well, from \cite{Ma17,Ma18}. We find a strong linear correlation (with Pearson correlation coefficients of $r=0.80$ and $p=0.02$) between $A_V$ and $W_\mathrm{r}(\lambda\,1560)$ for the GRB C\,\textsc{i} absorbers. We also note that C\,\textsc{i} is only detected in systems with $A_V \gtrsim 0.1$ mag, classifying the GRB C\,\textsc{i} systems as translucent interstellar clouds \citep[see e.g.][]{Snow06}. 

  \begin{figure*} %[!ht]
  	\centering
  	\epsfig{file=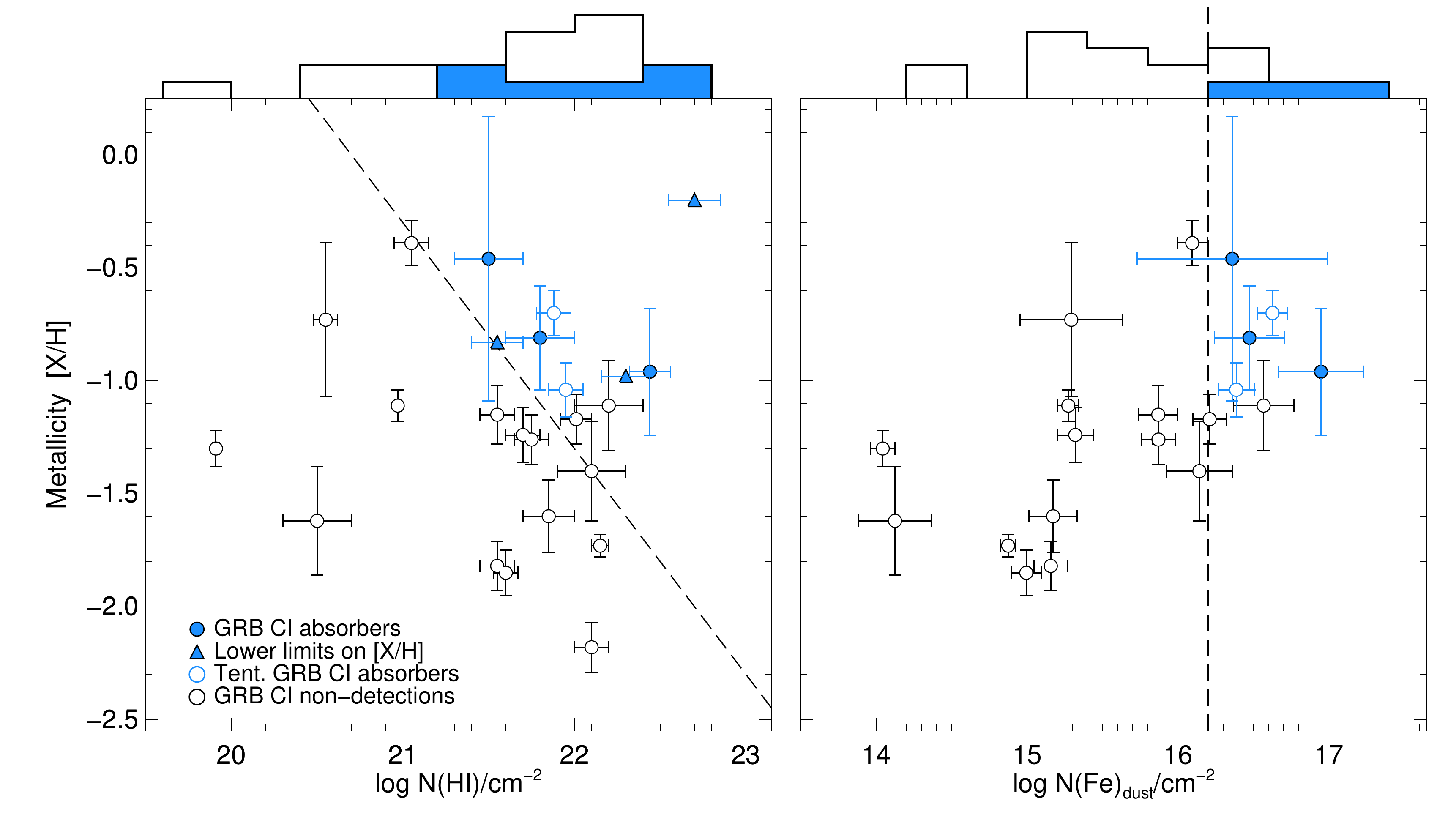,width=17cm}
  	\caption{Metallicity distribution as a function of H\,\textsc{i} column density (left panel) and dust-phase iron column density, $N$(Fe)$_{\mathrm{dust}}$ (right panel). Only the subsets of the full GRB sample with both a metallicity and $N$(H\,\textsc{i}) measurements (left) and depletion values (right) are shown. Blue filled circles again denote the GRBs with C\,\textsc{i} detected in absorption (empty blue circles show the GRBs with C\,\textsc{i} detected below the completeness limit). Black empty circles mark the GRBs in our samples without C\,\textsc{i}, where triangles represent bursts with lower limits on [X/H]. In the left panel, the dashed line corresponds to $\log N$(H\,\textsc{i})$/\mathrm{cm}^{-2}$ + [X/H] = 20.7, above which all the observed systems with detected C\,\textsc{i} are located and in the right panel the dashed line mark $\log N$(Fe)$_{\mathrm{dust}}/\mathrm{cm}^{-2} = 16.2$. At the top of both panels there are histograms of H\,\textsc{i} column density (left) and dust-phase iron column density (right) of the GRB absorbers with (filled blue) and without (black line) C\,\textsc{i} detected in absorption.}
  	\label{fig:metnhi}
  \end{figure*}   

In the right panel of Fig.~\ref{fig:w1560av} we again show the C\,\textsc{i} equivalent widths but as a function of the strength of the 2175\,\AA~dust extinction feature, $A_{\mathrm{bump}}$, defined as $A_{\mathrm{bump}} = \pi\,c_3 / (2\,\gamma\,R_V) \times A_V$. Only in five GRB absorbers to date has the 2175\,\AA~dust extinction feature been unambiguously observed \citep[][]{Kruehler08,Eliasdottir09,Prochaska09,Perley11,Zafar12,Zafar18b}, where C\,\textsc{i} is detected in four of them \citep[the non-detection in the last afterglow, GRB\,080805, is poorly constrained due to low S/N, however, see][]{Zafar12}. Here, we plot the four GRB C\,\textsc{i} absorbers with a clear presence of the 2175\,\AA~dust extinction feature (see Table~\ref{tab:obs}), where we have extracted the bump parameters for the four cases from the literature \citep[][]{Zafar11a,Zafar18b} and thus extend the analysis of \cite{Zafar12}. We also include the upper limits on the bump strength as measured for the other GRB C\,\textsc{i} absorber assuming the smooth SMC bar prescription by \cite{Gordon03}. We find evidence for a positive correlation between the amount of neutral carbon and the bump strength for the GRB C\,\textsc{i} absorbers. With only four detections of the 2175\,\AA~dust extinction feature in GRB C\,\textsc{i} absorbers, however, we are still limited by low number statistics. This relation has already been established by \cite{Ledoux15} and \cite{Ma18} for a larger number of quasar C\,\textsc{i} absorbers, but the addition of the small sample of GRB C\,\textsc{i} systems support this correlation and verify it at $A_{\mathrm{bump}} > 1$ mag as well.

The fact that C\,\textsc{i} absorbers are associated with significant dust columns is now firmly established \citep{Ledoux15,Ma18}. This could be related to larger dust columns more effectively shielding the neutral carbon or simply that C\,\textsc{i} scales with the overall amount of carbon which is dominant in the dust-phase. Nevertheless, it is established that in the cold neutral gas-phase, the amount of shielded gas and dust are connected. We note that among the quasar C\,\textsc{i} absorbers, some systems have large $W_\mathrm{r}(\lambda\,1560)$ but low extinction values which might indicate that these absorbers are metal-rich but with low gas content. The fact that no quasar C\,\textsc{i} absorbers with $A_V > 1$ mag have been detected might be due to a more severe dust bias in quasar DLA samples than for GRB afterglows \citep[e.g.][]{Heintz18b}. However, it could also simply be related to the fact that quasar DLAs are rarely observed with large $N$(H\,\textsc{i}) whereas GRB absorbers typically probe dense, central-galactic environments. We also find that the 2175\,\AA~dust extinction feature is only detected in GRBs with significant extinctions of $A_V > 0.5$ mag and strong C\,\textsc{i} absorption of $W_{\mathrm{r}}(\lambda\,1560) > 0.6$\,\AA. 
 
In Fig.~\ref{fig:hiav} we show the relation between $A_V$ and the column density of neutral hydrogen (H\,\textsc{i}) for the GRBs at $z\gtrsim 1.7$ in our sample and again compare them to the quasar C\,\textsc{i} absorbers from \cite{Ledoux15} and the 2175\,\AA~dust extinction feature absorbers from \cite{Ma17,Ma18}. For reference, the average dust-to-gas ratios from specific Galactic sightlines and toward the SMC bar, the mean LMC and the LMC2 supershell from \cite{Gordon03} are shown as well. In general, the GRB C\,\textsc{i} systems (and also the general sample of GRB absorbers) seem to have lower dust-to-gas ratios than the average Milky Way sightline, with a median value of $A_V/N$(H\,\textsc{i}) = $7.92\times 10^{-23}$ mag cm$^{-2}$. In the different regimes of H\,\textsc{i} covered by GRBs and the quasar absorbers, it is evident that the dust-to-gas ratio is significantly lower for the GRB C\,\textsc{i} systems than what is observed for the quasar C\,\textsc{i} absorbers. This is likely a consequence of the on average lower metallicities of the GRB host absorption systems \citep{Zafar11a} and the fact that the dust-to-metals ratio decrease with decreasing metallicity \citep[e.g.][]{DeCia13,Wiseman17}.

For the majority of the GRBs at $z\gtrsim 1.7$, we also obtained values for the metallicity, [X/H], from the literature (see Table~\ref{tab:obs}). This, together with the H\,\textsc{i} column density, yields the metal column density of the GRB absorbers, related to the dust column density \citep{Vladilo06,Zafar13}. In the left panel of Fig.~\ref{fig:metnhi} we show the metallicity as a function of H\,\textsc{i} column density for the GRBs in our parent sample. While the full sample spans a large range of 18.7 < $\log N$(H\,\textsc{i})$/\mathrm{cm}^{-2}$ < 22.7 and -2.2 < [X/H] < -0.2, it is clear that the GRB C\,\textsc{i} absorbers are all located above the threshold $\log N$(H\,\textsc{i})$/\mathrm{cm}^{-2}$ + [X/H] > 20.7 (marked as the dashed line). \cite{Bolmer18b} also found consistent results for H$_2$-bearing GRB absorbers, though extended to lower metallicities. We note that in the study of \cite{Ledoux09}, none of the GRBs in their sample have $\log N$(H\,\textsc{i})$/\mathrm{cm}^{-2}$ + [X/H] > 20.7. This threshold discovered here would then explain the non-detections of H$_2$ and C\,\textsc{i} absorption features in their high-resolution sample \citep[as speculated;][]{Ledoux09,Kruehler13}. 

To investigate more directly the influence of dust on the detection of C\,\textsc{i}, we derive the column density of iron locked into the dust-phase, defined as $N$(Fe)$_{\mathrm{dust}} = (1-10^{\mathrm{-[X/Fe]}})\,N(\mathrm{X})\,(\mathrm{Fe/X})_{\odot}$ \citep[e.g.][]{Vladilo06}. This quantity has been identified as the primary driver of the detection of H$_2$ molecules in quasar absorbers \citep{Noterdaeme08}. Only for a subset of the GRB afterglows in our sample at $z\gtrsim 1.7$, has the dust depletion, [X/Fe], been measured. Here it is assumed that X is a non-refractory element \citep[typically Zn or S, but see e.g.][]{Jenkins09,DeCia16} and that the intrinsic ratio is Solar. In the right panel of Fig.~\ref{fig:metnhi} we again show the metallicity but as a function of $N$(Fe)$_{\mathrm{dust}}$ for a subset of the GRB afterglow sample. We again observe a clear detection threshold of $\log N$(Fe)$_{\mathrm{dust}}/\mathrm{cm}^{-2} > 16.2$, above which all the GRB C\,\textsc{i} absorbers are located. 

We find no correlation between the metal or dust column density and the amount of neutral carbon. That is, a certain amount of metals is necessary to form dust and characterize absorbers with detected C\,\textsc{i}. However, the strength of $W_{\mathrm{r}}(\lambda\,1560)$ is found to be positively correlated only with the dust-reddening, $A_V$. Moreover, since the GRB C\,\textsc{i} absorbers are linked to the 2175\,\AA~dust extinction feature and also exclusively found at high $N$(Fe)$_{\mathrm{dust}}$, they must trace both carbon- and iron-rich dust, but the dust composition is likely predominantly carbon-rich since the amount of visual extinction, $A_V$, and the detection probability of the 2175\,\AA~dust extinction feature are found to be correlated with the amount of neutral carbon. 

Above the detection thresholds of $\log N$(H\,\textsc{i})$/\mathrm{cm}^{-2}$ + [X/H] > 20.7 and $\log N$(Fe)$_{\mathrm{dust}}/\mathrm{cm}^{-2} > 16.2$ two more GRBs (120815A and 121024A) are also observed with C\,\textsc{i} detected below the completeness limits. Two other bursts (GRBs\,120327A and 151021A) are observed in this region as well but without C\,\textsc{i} detected in absorption down to limits of $W_{\mathrm{r}}(\lambda\,1560) < 0.03$\,\AA\ and $W_{\mathrm{r}}(\lambda\,1560) < 0.17$\,\AA, respectively. GRB\,120327A has a reported detection of H$_2$, although with a low molecular fraction and dust extinction \citep{Delia14}.

%%%%%%%%%%%%%%%%%%%%%%%%%%%%%%%%%%%%%%%%%%%%%%%%%%%%%%%%%%%%%%%%%%%%%%%%%%%%%
\section{The diversity of cold gas absorbers} \label{sec:disc}
%%%%%%%%%%%%%%%%%%%%%%%%%%%%%%%%%%%%%%%%%%%%%%%%%%%%%%%%%%%%%%%%%%%%%%%%%%%%%

\begin{figure} [!t]
	\centering
	\epsfig{file=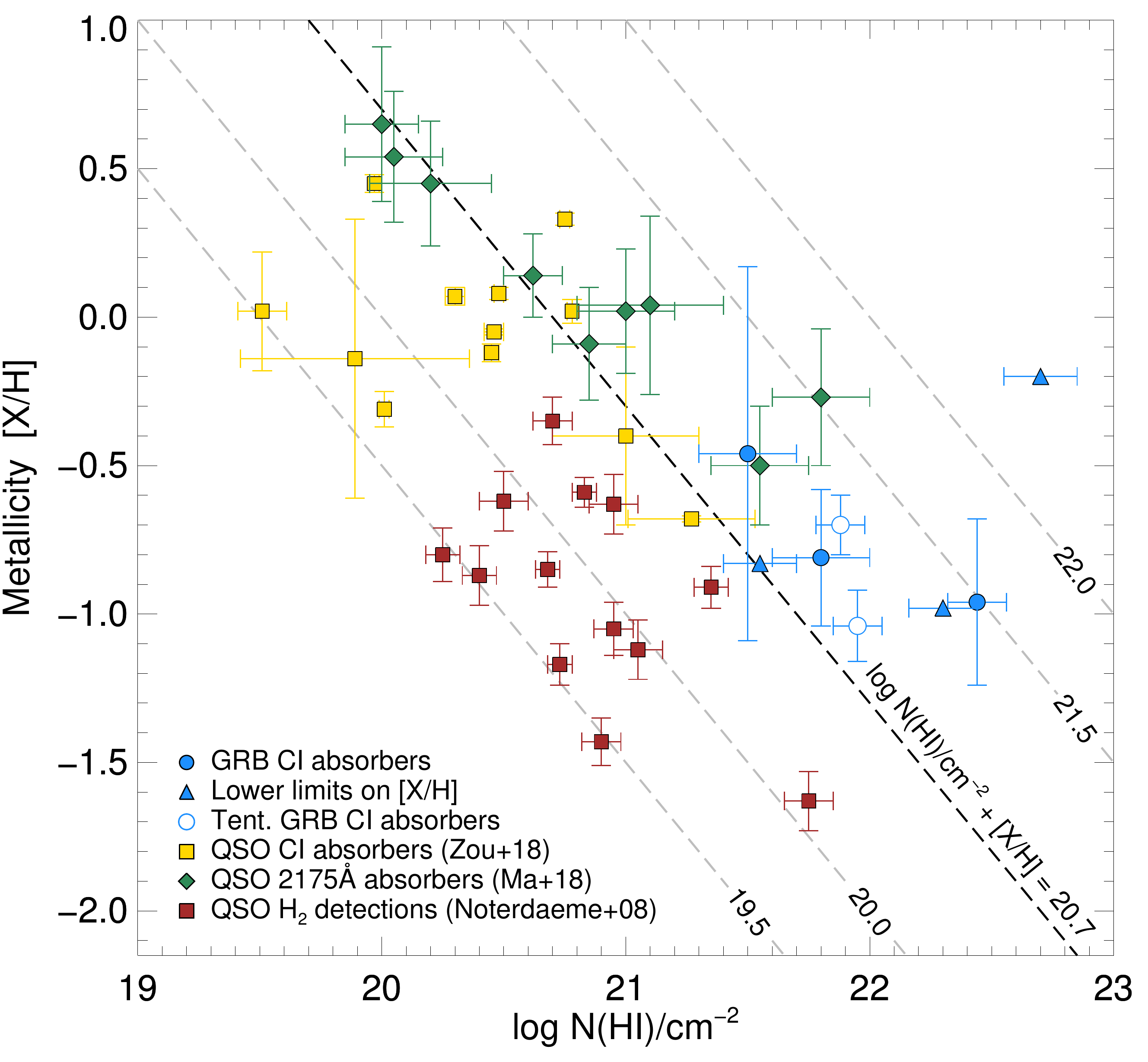,width=9cm,height=8.3cm}
	\caption{Comparison of the metallicity distribution as a function of H\,\textsc{i} column density for quasar and GRB absorbers with C\,\textsc{i} or H$_2$ detections. Blue filled circles again denote the GRBs with C\,\textsc{i} detected in absorption (empty blue circles show the GRBs with C\,\textsc{i} detected below the completeness limit). Overplotted are the quasar absorbers with H$_2$ detections from the VLT/UVES sample (dark red squares), quasar absorbers selected for the presence of the 2175\,\AA~dust extinction feature (green diamonds) and quasar absorbers selected for the presence of C\,\textsc{i} absorption (yellow squares). Only systems with $W_{\mathrm{r}}(\lambda\,1560) > 0.2$\,\AA\ from each of the C\,\textsc{i} samples are shown. The dashed lines mark constant values of $\log N$(H\,\textsc{i})$/\mathrm{cm}^{-2}$ + [X/H] as indicated at each line.}
	\label{fig:metnhicomp}
\end{figure}

\begin{figure} [!t]
	\centering
	\epsfig{file=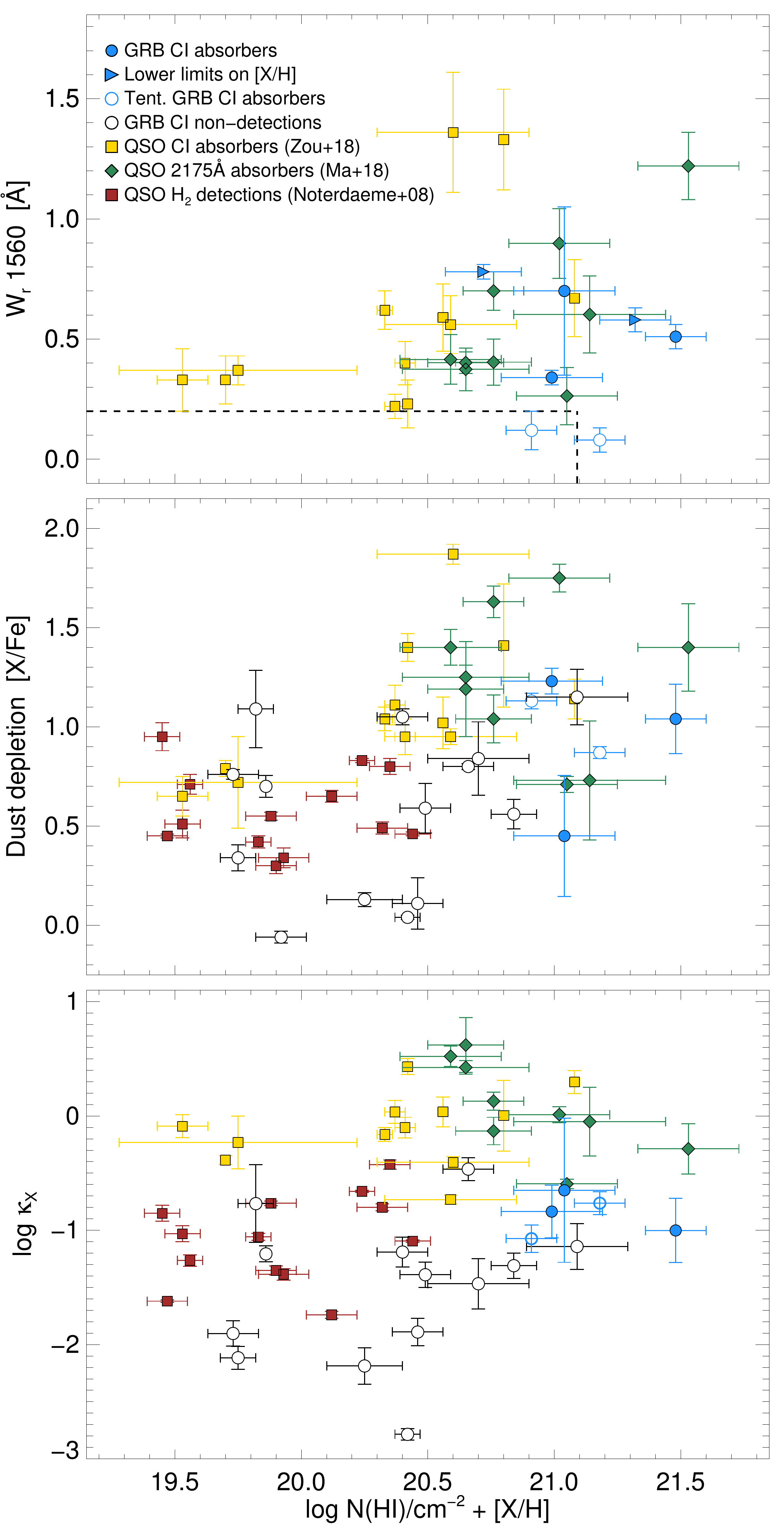,width=\columnwidth}
	\caption{$\log N$(H\,\textsc{i})$/\mathrm{cm}^{-2}$ + [X/H], as a function of $W_{\mathrm{r}}(\lambda\,1560)$ (top), dust depletion, [X/Fe], (middle) and depletion-derived dust-to-gas ratio, $\kappa_X$, (bottom) for quasar and GRB absorbers with C\,\textsc{i} or H$_2$ detections. The same symbols are used as in Fig.~\ref{fig:metnhicomp}, but here we also include the GRB afterglows with non-detections of C\,\textsc{i} (marked by the dashed lines in the top panel and as black empty circles in the bottom panels).}
	\label{fig:avdepldust}
\end{figure}

The presence of cold and molecular gas is regulated by the amount of dust in the ISM, that both serves to enhance the formation of molecular hydrogen onto dust grains but also lower the photo-dissociation rate by dust- and self-shielding. The detection probability of H$_2$ in quasar absorption systems is specifically found to be a function of metallicity and/or dust depletion \citep{Ledoux03,Srianand05,Petitjean06,Noterdaeme08}. In Fig.~\ref{fig:metnhicomp} we again plot the metallicity as a function of H\,\textsc{i} column density (similar to Fig.~\ref{fig:metnhi}) but now we focus on various types of absorption systems in which C\,\textsc{i} or H$_2$ (or both) have been detected. This is to investigate the diversity of the different types of these cold gas absorbers. We include the GRB C\,\textsc{i} absorbers from this study, the quasar C\,\textsc{i} absorbers from \cite{Ledoux15} with existing metallicities from \cite{Zou18}, and the quasar absorbers selected on the basis of the 2175\,\AA~dust extinction feature, all showing the presence of C\,\textsc{i}, from \cite{Ma17,Ma18}. For the comparison, we only include the systems with $W_{\mathrm{r}}(\lambda\,1560) > 0.2$\,\AA\ from the quasar absorber samples to match our completeness limit. We also include the quasar absorbers with H$_2$ detections from the high-resolution VLT/UVES sample by \cite{Noterdaeme08}.

We observe a notable difference between the GRB and quasar absorber populations: quasar absorbers with C\,\textsc{i} or H$_2$ detected in absorption are observed at significantly lower metal column densities than that of the GRB C\,\textsc{i} absorbers. In Fig.~\ref{fig:metnhicomp} we mark the lower limit in the H\,\textsc{i}-metallicity plane for which C\,\textsc{i} and H$_2$ has been detected in quasar absorbers as the bottom dashed line (at $\log N$(H\,\textsc{i})$/\mathrm{cm}^{-2}$ + [X/H] $\approx 19.5$). As can be seen from the figure, the cold neutral gas-phase is only observed in GRB hosts with metal column density $\sim 10$ times larger than typically observed for quasar absorbers selected on the basis of C\,\textsc{i} within the same completeness limit of $W_{\mathrm{r}}(\lambda\,1560) > 0.2$\,\AA. We caution that Fig.~\ref{fig:metnhicomp} mainly demonstrates the various selections from which these cold gas absorbers have been identified. GRB sightlines are expected to probe systematically higher metal column densities than quasar absorbers and the reason for the GRB C\,\textsc{i} absorbers to not be detected at lower metal column densities could simply be related to the small sample size not properly sampling the lower end of the parameter space. We emphasize, however, that the GRB afterglows observed with UVES as part of the F09 sample have $\log N$(H\,\textsc{i})$/\mathrm{cm}^{-2}$ + [X/H] in the same range ($\log N$(H\,\textsc{i})$/\mathrm{cm}^{-2}$ + [X/H] $= 19.7 - 20.7$), but show non-detections of C\,\textsc{i} (and H$_2$) down to similar deep limits as derived for the quasar H$_2$ absorbers. The non-detections in the GRB UVES sample could also be a consequence of the low sample size. However, while the general fraction of absorbers with cold gas in quasar sightlines is small \citep[5 -- 10\% estimated for H$_2$-bearing quasar absorbers;][]{Noterdaeme08,Balashev14,Balashev18}, the fraction increases significantly at $\log N$(H\,\textsc{i})$/\mathrm{cm}^{-2}$ + [X/H] $> 20$ \citep[to $\gtrsim 60\%$;][]{Noterdaeme15} and it would therefore be surprising that we do not detect C\,\textsc{i} in the GRB UVES sample solely due to the small sample size. A more plausible cause for the non-detection of C\,\textsc{i} in the GRB UVES sample is that it only consists of the brightest afterglows and therefore as a consequence is biased against dusty and metal-rich GRBs \citep{Ledoux09}.

The quasar absorbers selected on the basis of the 2175\,\AA~dust extinction feature appear mostly above the same threshold of $\log N$(H\,\textsc{i})$/\mathrm{cm}^{-2}$ + [X/H] > 20.7 as the GRB C\,\textsc{i} absorbers, though at higher metallicities and lower H\,\textsc{i} column densities. Among the quasar absorbers, the C\,\textsc{i}-selected systems appear to follow the same trend as the H$_2$-bearing absorbers, but are predominantly more metal-rich (as expected). On average, the GRB C\,\textsc{i} absorbers are the most gas-rich sightlines, with metal column densities comparable only to the quasar absorbers selected on the presence of the 2175\,\AA~dust extinction feature.  

In Fig.~\ref{fig:avdepldust} we compare the same samples examined above but study different dust indicators as a function of the metal column density. Here, we again include the GRBs in which C\,\textsc{i} is not detected. First, we show that the metal column density is not correlated with the amount of neutral carbon represented by $W_{\mathrm{r}}(\lambda\,1560)$ in any of the samples. Then, we investigate the detection probability of C\,\textsc{i} in GRB afterglows as a function of dust depletion, [X/Fe], in the middle panel of Fig.~\ref{fig:avdepldust}. This is motivated by the study of \cite{Ledoux03}, who found that the detection probability of H$_2$ in quasar absorbers is connected to the level of [X/Fe]. One burst, GRB\,070802, with C\,\textsc{i} detected in absorption, has a modest depletion value ([X/Fe] $=0.46$) and the remaining GRB C\,\textsc{i} systems have [X/Fe] $> 0.7$. However, there are several other GRBs without the presence of C\,\textsc{i} absorption features observed with similar high depletion values. This suggests that the presence of C\,\textsc{i} in GRB hosts is not particularly related to the amount of dust derived from the gas-phase depletion level. 
Following \cite{Ledoux03} we instead examine the dependence on the dust-to-gas ratio determined from the depletion level, described as $\kappa_{\mathrm{X}} = 10^{\mathrm{[X/H]}}(1-10^{\mathrm{-[X/Fe]}})$. They found that quasar H$_2$ absorbers were only detected at $\log\kappa_{\mathrm{X}} \gtrsim -1.5$, so this quantity could be more directly related to the presence of cold and molecular gas. Contrary to the dust depletion, we find a more distinct threshold for GRB C\,\textsc{i} absorbers to appear only above a certain depletion-derived dust-to-gas ratio of $\log\kappa_{\mathrm{X}} \gtrsim -1$ (see the bottom panel of Fig.~\ref{fig:avdepldust}). Only two of the GRB absorbers with non-detections of C\,\textsc{i} (GRBs\,050820A and 141028A) show the same depletion-derived dust-to-gas ratios. We note from Fig.~\ref{fig:avdepldust}, that $\kappa_{\mathrm{X}}$ is higher on average for the quasar H$_2$ absorbers compared to the GRBs with non-detections of C\,\textsc{i}. This could possibly explain why cold gas is detected in these particular quasar absorbers and not in the GRB absorbers with comparable metal column densities.

We therefore argue that the detection threshold of C\,\textsc{i} is primarily a function of the metal and dust column density. The fact that the cold neutral gas-phase in GRB hosts is only observed at metal and dust column densities more than an order of magnitude larger than for typical quasar absorbers is likely due to how these two absorber populations are selected. However, it could also be evidence of denser environments or more intense UV fields in the ISM of the star-forming GRB host galaxies \citep[as speculated;][]{Tumlinson07,Whalen08}. Finally, we note from Fig.~\ref{fig:metnhicomp} that C\,\textsc{i} or H$_2$-bearing quasar absorbers with high metal column densities are scarse. This deficit is likely not physical, however, but instead related to selection bias in quasar samples, systematically evading foreground absorbers rich in metals, dust and molecules \citep{Krogager16a,Fynbo17,Heintz18b}. Such a bias might also exist to a lesser extent in GRB afterglow samples. This bias will not affect the results of this paper, however, only underestimate the underlying population of GRB C\,\textsc{i} absorbers.

%%%%%%%%%%%%%%%%%%%%%%%%%%%%%%%%%%%%%%%%%%%%%%%%%%%%%%%%%%%%%%%%%%%%%%%%%%%%
\section{Conclusions} \label{sec:conc}
%%%%%%%%%%%%%%%%%%%%%%%%%%%%%%%%%%%%%%%%%%%%%%%%%%%%%%%%%%%%%%%%%%%%%%%%%%%%

In this work we presented a survey for neutral atomic-carbon (C\,\textsc{i}) in a sample of 29 medium- to high-resolution GRB afterglow spectra. We detected the absorption features of the C\,\textsc{i}\,$\lambda\lambda$\,1560,1656 line transitions in seven bursts ($\approx 25\%$) in our statistical sample at redshifts ranging from $1 < z < 4$. These GRB C\,\textsc{i} absorbers probe the shielded cold, neutral gas-phase of the ISM in their host galaxies and previous studies found the detection of C\,\textsc{i} to be directly linked to the presence of molecular hydrogen \citep{Srianand05,Noterdaeme18}.

Our goals were to characterize the dust properties (such as the amount of extinction and the strength of the 2175\,\AA~dust extinction feature) and the chemical abundances of these GRB C\,\textsc{i} absorbers. We found that the amount of neutral carbon is positively correlated with the visual extinction, $A_V$, and the strength of the 2175\,\AA~dust extinction feature \citep[as also observed for quasar C\,\textsc{i} absorbers;][]{Ledoux15,Ma18}. These relations support a scenario where it is predominantly carbonaceous dust grains that produce the characteristic bump \citep{Henning98,Draine03}. The average dust-to-gas ratio of the GRB C\,\textsc{i} absorbers was found to be significantly smaller than observed for quasar C\,\textsc{i} absorbers and in typical Milky Way sightlines, with a median value of $A_V/N$(H\,\textsc{i}) = $7.92\times 10^{-23}$\,mag\,cm$^{-2}$. We showed that C\,\textsc{i} is only observed in GRB host absorption systems above a certain threshold of $\log N$(H\,\textsc{i})$/\mathrm{cm}^{-2}$ + [X/H] and dust-phase iron column densities of $\log N$(Fe)$_{\mathrm{dust}}/\mathrm{cm}^{-2} > 16.2$. The connection of the GRB C\,\textsc{i} systems to the 2175\,\AA~dust extinction feature and the large values of $N$(Fe)$_{\mathrm{dust}}$ indicate that the C\,\textsc{i} absorbers trace dusty systems with a dust composition that both consist of carbon- and iron-rich dust grains, but is dominated by the carbon-rich dust.

The fact that the metal and dust column densities of the GRB C\,\textsc{i} absorbers are higher than observed for C\,\textsc{i}- and H$_2$-bearing quasar absorbers is primarily a consequence of how these two absorber populations are selected, but is also consistent with a scenario where GRB hosts have more intense galactic UV fields than the galaxy counterparts of the quasar absorbers. We also investigated the detection probability of C\,\textsc{i} as a function of other dust tracers such as the gas-phase depletion, [X/Fe], and the depletion-derived dust-to-gas ratio, $\kappa_\mathrm{X}$. We found that the metal and dust column densities are the primary driver for the presence of C\,\textsc{i} in GRB hosts, although the amount of neutral carbon is not correlated with either. Instead, the amount of neutral carbon is positively correlated with the visual extinction, $A_V$, but is detected down to relatively small visual extinctions of $A_V \approx 0.1$ mag.

We argue that C\,\textsc{i} has a number of advantages as a tracer of the cold neutral medium in GRB hosts due to the observational limitations of identifying H$_2$ absorption features in the majority of GRB afterglow spectra. A more detailed analysis of the relations between H$_2$, C\,\textsc{i} and CO column densities in GRB absorbers is needed to better constrain the physical properties of the cold gas. Follow-up observations of the host galaxies of the GRB C\,\textsc{i} absorbers identified here at millimetre wavelengths with e.g. ALMA would also be interesting to pursue in the future.

%%%%%%%%%%%%%%%%%%%%%%%%%%%%%%%%%%%%%%%%%%%%%%%%%%%%%%%%%%%%%%%%%%%%%%%%%%%%

\begin{acknowledgements}
We would like to thank the referee for a clear, constructive and concise report that greatly improved the presentation of the results from this work.
We also wish to thank the large European GRB collaboration and the dedicated ToO advocates staying on alert at all times, day and night. Without the swift reaction of this team, such a large number of good quality GRB afterglow spectra would never have been gathered. KEH and PJ acknowledge support by a Project Grant (162948--051) from The Icelandic Research Fund. The Cosmic Dawn Center is funded by the DNRF. JK and PN acknowledge funding from the French {\sl Agence Nationale de la Recherche} under grant no ANR-17-CE31-0011-01. SDV acknowledges the support of the French National Research Agency (ANR) under contract ANR-16-CE31-0003. JJ acknowledges support from NOVA and NWO-FAPESP grant for advanced instrumentation in astronomy.
\end{acknowledgements}

\bibliographystyle{aa}
\bibliography{ref}

\begin{thebibliography}{107}
\expandafter\ifx\csname natexlab\endcsname\relax\def\natexlab#1{#1}\fi

\bibitem[{{Amanullah} {et~al.}(2014){Amanullah}, {Goobar}, {Johansson},
  {Banerjee}, {Venkataraman}, {Joshi}, {Ashok}, {Cao}, {Kasliwal}, {Kulkarni},
  {Nugent}, {Petrushevska}, \& {Stanishev}}]{Amanullah14}
{Amanullah}, R., {Goobar}, A., {Johansson}, J., {et~al.} 2014, \apjl, 788, L21

\bibitem[{{Asplund} {et~al.}(2009){Asplund}, {Grevesse}, {Sauval}, \&
  {Scott}}]{Asplund09}
{Asplund}, M., {Grevesse}, N., {Sauval}, A.~J., \& {Scott}, P. 2009, \araa, 47,
  481

\bibitem[{{Atwood} {et~al.}(2009){Atwood}, {Abdo}, {Ackermann}, {Althouse},
  {Anderson}, {Axelsson}, {Baldini}, {Ballet}, {Band}, {Barbiellini}, \&
  et~al.}]{Atwood09}
{Atwood}, W.~B., {Abdo}, A.~A., {Ackermann}, M., {et~al.} 2009, \apj, 697, 1071

\bibitem[{{Balashev} {et~al.}(2014){Balashev}, {Klimenko}, {Ivanchik},
  {Varshalovich}, {Petitjean}, \& {Noterdaeme}}]{Balashev14}
{Balashev}, S.~A., {Klimenko}, V.~V., {Ivanchik}, A.~V., {et~al.} 2014, \mnras,
  440, 225

\bibitem[{{Balashev} \& {Noterdaeme}(2018)}]{Balashev18}
{Balashev}, S.~A. \& {Noterdaeme}, P. 2018, \mnras, 478, L7

\bibitem[{{Boiss\'e} {et~al.}(1998){Boiss\'e}, {Le Brun}, {Bergeron}, \&
  {Deharveng}}]{Boisse98}
{Boiss\'e}, P., {Le Brun}, V., {Bergeron}, J., \& {Deharveng}, J.-M. 1998,
  \aap, 333, 841

\bibitem[{{Bolmer} {et~al.}(2018{\natexlab{a}}){Bolmer}, {Greiner},
  {Kr{\"u}hler}, {Schady}, {Ledoux}, {Tanvir}, \& {Levan}}]{Bolmer18a}
{Bolmer}, J., {Greiner}, J., {Kr{\"u}hler}, T., {et~al.} 2018{\natexlab{a}},
  \aap, 609, A62

\bibitem[{{Bolmer} {et~al.}(2018{\natexlab{b}}){Bolmer}, {Ledoux}, {Wiseman},
  {De Cia}, {Selsing}, {Schady}, {Greiner}, {Savaglio}, {Burgess}, {D'Elia},
  {Fynbo}, {Goldoni}, {Hartmann}, {Heintz}, {Jakobsson}, {Japelj}, {Kaper},
  {Tanvir}, {Vreeswijk}, \& {Zafar}}]{Bolmer18b}
{Bolmer}, J., {Ledoux}, C., {Wiseman}, P., {et~al.} 2018{\natexlab{b}}, ArXiv
  (1810.06403)

\bibitem[{{Cameron}(2011)}]{Cameron11}
{Cameron}, E. 2011, \pasa, 28, 128

\bibitem[{{Christensen} {et~al.}(2004){Christensen}, {Hjorth}, \&
  {Gorosabel}}]{Christensen04}
{Christensen}, L., {Hjorth}, J., \& {Gorosabel}, J. 2004, \aap, 425, 913

\bibitem[{{Covino} {et~al.}(2013){Covino}, {Melandri}, {Salvaterra}, {Campana},
  {Vergani}, {Bernardini}, {D'Avanzo}, {D'Elia}, {Fugazza}, {Ghirlanda},
  {Ghisellini}, {Gomboc}, {Jin}, {Kr{\"u}hler}, {Malesani}, {Nava},
  {Sbarufatti}, \& {Tagliaferri}}]{Covino13}
{Covino}, S., {Melandri}, A., {Salvaterra}, R., {et~al.} 2013, \mnras, 432,
  1231

\bibitem[{{Cucchiara} {et~al.}(2015){Cucchiara}, {Fumagalli}, {Rafelski},
  {Kocevski}, {Prochaska}, {Cooke}, \& {Becker}}]{Cucchiara15}
{Cucchiara}, A., {Fumagalli}, M., {Rafelski}, M., {et~al.} 2015, \apj, 804, 51

\bibitem[{{Cucchiara} {et~al.}(2011){Cucchiara}, {Levan}, {Fox}, {Tanvir},
  {Ukwatta}, {Berger}, {Kr{\"u}hler}, {K{\"u}pc{\"u} Yolda{\c s}}, {Wu},
  {Toma}, {Greiner}, {Olivares}, {Rowlinson}, {Amati}, {Sakamoto}, {Roth},
  {Stephens}, {Fritz}, {Fynbo}, {Hjorth}, {Malesani}, {Jakobsson}, {Wiersema},
  {O'Brien}, {Soderberg}, {Foley}, {Fruchter}, {Rhoads}, {Rutledge}, {Schmidt},
  {Dopita}, {Podsiadlowski}, {Willingale}, {Wolf}, {Kulkarni}, \&
  {D'Avanzo}}]{Cucchiara11}
{Cucchiara}, A., {Levan}, A.~J., {Fox}, D.~B., {et~al.} 2011, \apj, 736, 7

\bibitem[{{De Cia} {et~al.}(2012){De Cia}, {Ledoux}, {Fox}, {Vreeswijk},
  {Smette}, {Petitjean}, {Bj{\"o}rnsson}, {Fynbo}, {Hjorth}, \&
  {Jakobsson}}]{DeCia12}
{De Cia}, A., {Ledoux}, C., {Fox}, A.~J., {et~al.} 2012, \aap, 545, A64

\bibitem[{{De Cia} {et~al.}(2016){De Cia}, {Ledoux}, {Mattsson}, {Petitjean},
  {Srianand}, {Gavignaud}, \& {Jenkins}}]{DeCia16}
{De Cia}, A., {Ledoux}, C., {Mattsson}, L., {et~al.} 2016, \aap, 596, A97

\bibitem[{{De Cia} {et~al.}(2013){De Cia}, {Ledoux}, {Savaglio}, {Schady}, \&
  {Vreeswijk}}]{DeCia13}
{De Cia}, A., {Ledoux}, C., {Savaglio}, S., {Schady}, P., \& {Vreeswijk}, P.~M.
  2013, \aap, 560, A88

\bibitem[{{de Ugarte Postigo} {et~al.}(2018){de Ugarte Postigo}, {Th{\"o}ne},
  {Bolmer}, {Schulze}, {Mart{\'{\i}}n}, {Kann}, {D'Elia}, {Selsing},
  {Martin-Carrillo}, {Perley}, {Kim}, {Izzo}, {S{\'a}nchez-Ram{\'{\i}}rez},
  {Guidorzi}, {Klotz}, {Wiersema}, {Bauer}, {Bensch}, {Campana}, {Cano},
  {Covino}, {Coward}, {De Cia}, {de Gregorio-Monsalvo}, {De Pasquale}, {Fynbo},
  {Greiner}, {Gomboc}, {Hanlon}, {Hansen}, {Hartmann}, {Heintz}, {Jakobsson},
  {Kobayashi}, {Malesani}, {Martone}, {Meintjes}, {Michalowski}, {Mundell},
  {Murphy}, {Oates}, {Resmi}, {Salmon}, {van Soelen}, {Tanvir}, {Turpin}, {Xu},
  \& {Zafar}}]{DeUgartePostigo18}
{de Ugarte Postigo}, A., {Th{\"o}ne}, C.~C., {Bolmer}, J., {et~al.} 2018, ArXiv
  (1806.07393)

\bibitem[{{D'Elia} {et~al.}(2007){D'Elia}, {Fiore}, {Meurs}, {Chincarini},
  {Melandri}, {Norci}, {Pellizza}, {Perna}, {Piranomonte}, {Sbordone},
  {Stella}, {Tagliaferri}, {Vergani}, {Ward}, {Angelini}, {Antonelli},
  {Burrows}, {Campana}, {Capalbi}, {Cimatti}, {Costa}, {Cusumano}, {Della
  Valle}, {Filliatre}, {Fontana}, {Frontera}, {Fugazza}, {Gehrels}, {Giannini},
  {Giommi}, {Goldoni}, {Guetta}, {Israel}, {Lazzati}, {Malesani}, {Marconi},
  {Mason}, {Mereghetti}, {Mirabel}, {Molinari}, {Moretti}, {Nousek}, {Perri},
  {Piro}, {Stratta}, {Testa}, \& {Vietri}}]{Delia07}
{D'Elia}, V., {Fiore}, F., {Meurs}, E.~J.~A., {et~al.} 2007, \aap, 467, 629

\bibitem[{{D'Elia} {et~al.}(2010){D'Elia}, {Fynbo}, {Covino}, {Goldoni},
  {Jakobsson}, {Matteucci}, {Piranomonte}, {Sollerman}, {Th{\"o}ne}, {Vergani},
  {Vreeswijk}, {Watson}, {Wiersema}, {Zafar}, {de Ugarte Postigo}, {Flores},
  {Hjorth}, {Kaper}, {Levan}, {Malesani}, {Milvang-Jensen}, {Pian},
  {Tagliaferri}, \& {Tanvir}}]{Delia10}
{D'Elia}, V., {Fynbo}, J.~P.~U., {Covino}, S., {et~al.} 2010, \aap, 523, A36

\bibitem[{{D'Elia} {et~al.}(2014){D'Elia}, {Fynbo}, {Goldoni}, {Covino}, {de
  Ugarte Postigo}, {Ledoux}, {Calura}, {Gorosabel}, {Malesani}, {Matteucci},
  {S{\'a}nchez-Ram{\'{\i}}rez}, {Savaglio}, {Castro-Tirado}, {Hartoog},
  {Kaper}, {Mu{\~n}oz-Darias}, {Pian}, {Piranomonte}, {Tagliaferri}, {Tanvir},
  {Vergani}, {Watson}, \& {Xu}}]{Delia14}
{D'Elia}, V., {Fynbo}, J.~P.~U., {Goldoni}, P., {et~al.} 2014, \aap, 564, A38

\bibitem[{{Draine}(2003)}]{Draine03}
{Draine}, B.~T. 2003, \araa, 41, 241

\bibitem[{{Draine} \& {Hao}(2002)}]{Draine02}
{Draine}, B.~T. \& {Hao}, L. 2002, \apj, 569, 780

\bibitem[{{El{\'{\i}}asd{\'o}ttir} {et~al.}(2009){El{\'{\i}}asd{\'o}ttir},
  {Fynbo}, {Hjorth}, {Ledoux}, {Watson}, {Andersen}, {Malesani}, {Vreeswijk},
  {Prochaska}, {Sollerman}, \& {Jaunsen}}]{Eliasdottir09}
{El{\'{\i}}asd{\'o}ttir}, {\'A}., {Fynbo}, J.~P.~U., {Hjorth}, J., {et~al.}
  2009, \apj, 697, 1725

\bibitem[{{Fall} \& {Pei}(1993)}]{Fall93}
{Fall}, S.~M. \& {Pei}, Y.~C. 1993, \apj, 402, 479

\bibitem[{{Fitzpatrick} \& {Massa}(1990)}]{Fitzpatrick90}
{Fitzpatrick}, E.~L. \& {Massa}, D. 1990, \apjs, 72, 163

\bibitem[{{Friis} {et~al.}(2015){Friis}, {De Cia}, {Kr{\"u}hler}, {Fynbo},
  {Ledoux}, {Vreeswijk}, {Watson}, {Malesani}, {Gorosabel}, {Starling},
  {Jakobsson}, {Varela}, {Wiersema}, {Drachmann}, {Trotter}, {Th{\"o}ne}, {de
  Ugarte Postigo}, {D'Elia}, {Elliott}, {Maturi}, {Goldoni}, {Greiner},
  {Haislip}, {Kaper}, {Knust}, {LaCluyze}, {Milvang-Jensen}, {Reichart},
  {Schulze}, {Sudilovsky}, {Tanvir}, \& {Vergani}}]{Friis15}
{Friis}, M., {De Cia}, A., {Kr{\"u}hler}, T., {et~al.} 2015, \mnras, 451, 167

\bibitem[{{Fynbo} {et~al.}(2009){Fynbo}, {Jakobsson}, {Prochaska}, {Malesani},
  {Ledoux}, {de Ugarte Postigo}, {Nardini}, {Vreeswijk}, {Wiersema}, {Hjorth},
  {Sollerman}, {Chen}, {Th{\"o}ne}, {Bj{\"o}rnsson}, {Bloom}, {Castro-Tirado},
  {Christensen}, {De Cia}, {Fruchter}, {Gorosabel}, {Graham}, {Jaunsen},
  {Jensen}, {Kann}, {Kouveliotou}, {Levan}, {Maund}, {Masetti},
  {Milvang-Jensen}, {Palazzi}, {Perley}, {Pian}, {Rol}, {Schady}, {Starling},
  {Tanvir}, {Watson}, {Xu}, {Augusteijn}, {Grundahl}, {Telting}, \&
  {Quirion}}]{Fynbo09}
{Fynbo}, J.~P.~U., {Jakobsson}, P., {Prochaska}, J.~X., {et~al.} 2009, \apjs,
  185, 526

\bibitem[{{Fynbo} {et~al.}(2017){Fynbo}, {Krogager}, {Heintz}, {Geier},
  {M{\o}ller}, {Noterdaeme}, {Christensen}, {Ledoux}, \& {Jakobsson}}]{Fynbo17}
{Fynbo}, J.~P.~U., {Krogager}, J.-K., {Heintz}, K.~E., {et~al.} 2017, \aap,
  606, A13

\bibitem[{{Fynbo} {et~al.}(2013){Fynbo}, {Krogager}, {Venemans}, {Noterdaeme},
  {Vestergaard}, {M{\o}ller}, {Ledoux}, \& {Geier}}]{Fynbo13}
{Fynbo}, J.~P.~U., {Krogager}, J.-K., {Venemans}, B., {et~al.} 2013, \apjs,
  204, 6

\bibitem[{{Fynbo} {et~al.}(2014){Fynbo}, {Kr{\"u}hler}, {Leighly}, {Ledoux},
  {Vreeswijk}, {Schulze}, {Noterdaeme}, {Watson}, {Wijers}, {Bolmer}, {Cano},
  {Christensen}, {Covino}, {D'Elia}, {Flores}, {Friis}, {Goldoni}, {Greiner},
  {Hammer}, {Hjorth}, {Jakobsson}, {Japelj}, {Kaper}, {Klose}, {Knust},
  {Leloudas}, {Levan}, {Malesani}, {Milvang-Jensen}, {M{\o}ller}, {Nicuesa
  Guelbenzu}, {Oates}, {Pian}, {Schady}, {Sparre}, {Tagliaferri}, {Tanvir},
  {Th{\"o}ne}, {de Ugarte Postigo}, {Vergani}, {Wiersema}, {Xu}, \&
  {Zafar}}]{Fynbo14}
{Fynbo}, J.~P.~U., {Kr{\"u}hler}, T., {Leighly}, K., {et~al.} 2014, \aap, 572,
  A12

\bibitem[{{Fynbo} {et~al.}(2001){Fynbo}, {Jensen}, {Gorosabel}, {Hjorth},
  {Pedersen}, {M{\o}ller}, {Abbott}, {Castro-Tirado}, {Delgado}, {Greiner},
  {Henden}, {Magazz{\`u}}, {Masetti}, {Merlino}, {Masegosa}, {{\O}stensen},
  {Palazzi}, {Pian}, {Schwarz}, {Cline}, {Guidorzi}, {Goldsten}, {Hurley},
  {Mazets}, {McClanahan}, {Montanari}, {Starr}, \& {Trombka}}]{Fynbo01}
{Fynbo}, J.~U., {Jensen}, B.~L., {Gorosabel}, J., {et~al.} 2001, \aap, 369, 373

\bibitem[{{Gehrels} {et~al.}(2004){Gehrels}, {Chincarini}, {Giommi}, {Mason},
  {Nousek}, {Wells}, {White}, {Barthelmy}, {Burrows}, {Cominsky}, {Hurley},
  {Marshall}, {M{\'e}sz{\'a}ros}, {Roming}, {Angelini}, {Barbier}, {Belloni},
  {Campana}, {Caraveo}, {Chester}, {Citterio}, {Cline}, {Cropper}, {Cummings},
  {Dean}, {Feigelson}, {Fenimore}, {Frail}, {Fruchter}, {Garmire}, {Gendreau},
  {Ghisellini}, {Greiner}, {Hill}, {Hunsberger}, {Krimm}, {Kulkarni}, {Kumar},
  {Lebrun}, {Lloyd-Ronning}, {Markwardt}, {Mattson}, {Mushotzky}, {Norris},
  {Osborne}, {Paczynski}, {Palmer}, {Park}, {Parsons}, {Paul}, {Rees},
  {Reynolds}, {Rhoads}, {Sasseen}, {Schaefer}, {Short}, {Smale}, {Smith},
  {Stella}, {Tagliaferri}, {Takahashi}, {Tashiro}, {Townsley}, {Tueller},
  {Turner}, {Vietri}, {Voges}, {Ward}, {Willingale}, {Zerbi}, \&
  {Zhang}}]{Gehrels04}
{Gehrels}, N., {Chincarini}, G., {Giommi}, P., {et~al.} 2004, \apj, 611, 1005

\bibitem[{{Glover} \& {Clark}(2016)}]{Glover16}
{Glover}, S.~C.~O. \& {Clark}, P.~C. 2016, \mnras, 456, 3596

\bibitem[{{Gordon} {et~al.}(2003){Gordon}, {Clayton}, {Misselt}, {Landolt}, \&
  {Wolff}}]{Gordon03}
{Gordon}, K.~D., {Clayton}, G.~C., {Misselt}, K.~A., {Landolt}, A.~U., \&
  {Wolff}, M.~J. 2003, \apj, 594, 279

\bibitem[{{Greiner} {et~al.}(2015){Greiner}, {Fox}, {Schady}, {Kr{\"u}hler},
  {Trenti}, {Cikota}, {Bolmer}, {Elliott}, {Delvaux}, {Perna}, {Afonso},
  {Kann}, {Klose}, {Savaglio}, {Schmidl}, {Schweyer}, {Tanga}, \&
  {Varela}}]{Greiner15}
{Greiner}, J., {Fox}, D.~B., {Schady}, P., {et~al.} 2015, \apj, 809, 76

\bibitem[{{Greiner} {et~al.}(2011){Greiner}, {Kr{\"u}hler}, {Klose}, {Afonso},
  {Clemens}, {Filgas}, {Hartmann}, {K{\"u}pc{\"u} Yolda{\c s}}, {Nardini},
  {Olivares E.}, {Rau}, {Rossi}, {Schady}, \& {Updike}}]{Greiner11}
{Greiner}, J., {Kr{\"u}hler}, T., {Klose}, S., {et~al.} 2011, \aap, 526, A30

\bibitem[{{Hartoog} {et~al.}(2015){Hartoog}, {Malesani}, {Fynbo}, {Goto},
  {Kr{\"u}hler}, {Vreeswijk}, {De Cia}, {Xu}, {M{\o}ller}, {Covino}, {D'Elia},
  {Flores}, {Goldoni}, {Hjorth}, {Jakobsson}, {Krogager}, {Kaper}, {Ledoux},
  {Levan}, {Milvang-Jensen}, {Sollerman}, {Sparre}, {Tagliaferri}, {Tanvir},
  {de Ugarte Postigo}, {Vergani}, {Wiersema}, {Datson}, {Salinas}, {Mikkelsen},
  \& {Aghanim}}]{Hartoog15}
{Hartoog}, O.~E., {Malesani}, D., {Fynbo}, J.~P.~U., {et~al.} 2015, \aap, 580,
  A139

\bibitem[{{Heintz} {et~al.}(2017){Heintz}, {Fynbo}, {Jakobsson}, {Kr{\"u}hler},
  {Christensen}, {Watson}, {Ledoux}, {Noterdaeme}, {Perley}, {Rhodin},
  {Selsing}, {Schulze}, {Tanvir}, {M{\o}ller}, {Goldoni}, {Xu}, \&
  {Milvang-Jensen}}]{Heintz17}
{Heintz}, K.~E., {Fynbo}, J.~P.~U., {Jakobsson}, P., {et~al.} 2017, \aap, 601,
  A83

\bibitem[{{Heintz} {et~al.}(2018{\natexlab{a}}){Heintz}, {Fynbo}, {Ledoux},
  {Jakobsson}, {M{\o}ller}, {Christensen}, {Geier}, {Krogager}, \&
  {Noterdaeme}}]{Heintz18b}
{Heintz}, K.~E., {Fynbo}, J.~P.~U., {Ledoux}, C., {et~al.} 2018{\natexlab{a}},
  \aap, 615, A43

\bibitem[{{Heintz} {et~al.}(2018{\natexlab{b}}){Heintz}, {Watson}, {Jakobsson},
  {Fynbo}, {Bolmer}, {Arabsalmani}, {Cano}, {Covino}, {D'Elia}, {Gomboc},
  {Japelj}, {Kaper}, {Krogager}, {Pugliese}, {S{\'a}nchez-Ram{\'{\i}}rez},
  {Selsing}, {Sparre}, {Tanvir}, {Th{\"o}ne}, {de Ugarte Postigo}, \&
  {Vergani}}]{Heintz18a}
{Heintz}, K.~E., {Watson}, D., {Jakobsson}, P., {et~al.} 2018{\natexlab{b}},
  \mnras, 479, 3456

\bibitem[{{Henning} \& {Salama}(1998)}]{Henning98}
{Henning}, T. \& {Salama}, F. 1998, Science, 282, 2204

\bibitem[{{Jakobsson} {et~al.}(2005){Jakobsson}, {Bj{\"o}rnsson}, {Fynbo},
  {J{\'o}hannesson}, {Hjorth}, {Thomsen}, {M{\o}ller}, {Watson}, {Jensen},
  {{\"O}stlin}, {Gorosabel}, \& {Gudmundsson}}]{Jakobsson05}
{Jakobsson}, P., {Bj{\"o}rnsson}, G., {Fynbo}, J.~P.~U., {et~al.} 2005, \mnras,
  362, 245

\bibitem[{{Jakobsson} {et~al.}(2004){Jakobsson}, {Hjorth}, {Fynbo}, {Watson},
  {Pedersen}, {Bj{\"o}rnsson}, \& {Gorosabel}}]{Jakobsson04}
{Jakobsson}, P., {Hjorth}, J., {Fynbo}, J.~P.~U., {et~al.} 2004, \apjl, 617,
  L21

\bibitem[{{Japelj} {et~al.}(2015){Japelj}, {Covino}, {Gomboc}, {Vergani},
  {Goldoni}, {Selsing}, {Cano}, {D'Elia}, {Flores}, {Fynbo}, {Hammer},
  {Hjorth}, {Jakobsson}, {Kaper}, {Kopa{\v c}}, {Kr{\"u}hler}, {Melandri},
  {Piranomonte}, {S{\'a}nchez-Ram{\'{\i}}rez}, {Tagliaferri}, {Tanvir}, {de
  Ugarte Postigo}, {Watson}, \& {Wijers}}]{Japelj15}
{Japelj}, J., {Covino}, S., {Gomboc}, A., {et~al.} 2015, \aap, 579, A74

\bibitem[{{Jenkins}(2009)}]{Jenkins09}
{Jenkins}, E.~B. 2009, \apj, 700, 1299

\bibitem[{{Jiang} {et~al.}(2010){Jiang}, {Ge}, {Prochaska}, {Kulkarni}, {Lu},
  \& {Zhou}}]{Jiang10}
{Jiang}, P., {Ge}, J., {Prochaska}, J.~X., {et~al.} 2010, \apj, 720, 328

\bibitem[{{Kann} {et~al.}(2010){Kann}, {Klose}, {Zhang}, {Malesani}, {Nakar},
  {Pozanenko}, {Wilson}, {Butler}, {Jakobsson}, {Schulze}, {Andreev},
  {Antonelli}, {Bikmaev}, {Biryukov}, {B{\"o}ttcher}, {Burenin}, {Castro
  Cer{\'o}n}, {Castro-Tirado}, {Chincarini}, {Cobb}, {Covino}, {D'Avanzo},
  {D'Elia}, {Della Valle}, {de Ugarte Postigo}, {Efimov}, {Ferrero}, {Fugazza},
  {Fynbo}, {G{\aa}lfalk}, {Grundahl}, {Gorosabel}, {Gupta}, {Guziy}, {Hafizov},
  {Hjorth}, {Holhjem}, {Ibrahimov}, {Im}, {Israel}, {Je{\'l}inek}, {Jensen},
  {Karimov}, {Khamitov}, {Kizilo{\v g}lu}, {Klunko}, {Kub{\'a}nek}, {Kutyrev},
  {Laursen}, {Levan}, {Mannucci}, {Martin}, {Mescheryakov}, {Mirabal},
  {Norris}, {Ovaldsen}, {Paraficz}, {Pavlenko}, {Piranomonte}, {Rossi},
  {Rumyantsev}, {Salinas}, {Sergeev}, {Sharapov}, {Sollerman}, {Stecklum},
  {Stella}, {Tagliaferri}, {Tanvir}, {Telting}, {Testa}, {Updike}, {Volnova},
  {Watson}, {Wiersema}, \& {Xu}}]{Kann10}
{Kann}, D.~A., {Klose}, S., {Zhang}, B., {et~al.} 2010, \apj, 720, 1513

\bibitem[{{Kistler} {et~al.}(2009){Kistler}, {Y{\"u}ksel}, {Beacom}, {Hopkins},
  \& {Wyithe}}]{Kistler09}
{Kistler}, M.~D., {Y{\"u}ksel}, H., {Beacom}, J.~F., {Hopkins}, A.~M., \&
  {Wyithe}, J.~S.~B. 2009, \apjl, 705, L104

\bibitem[{{Krogager} {et~al.}(2016{\natexlab{a}}){Krogager}, {Fynbo}, {Heintz},
  {Geier}, {Ledoux}, {M{\o}ller}, {Noterdaeme}, {Venemans}, \&
  {Vestergaard}}]{Krogager16b}
{Krogager}, J.-K., {Fynbo}, J.~P.~U., {Heintz}, K.~E., {et~al.}
  2016{\natexlab{a}}, \apj, 832, 49

\bibitem[{{Krogager} {et~al.}(2016{\natexlab{b}}){Krogager}, {Fynbo},
  {Noterdaeme}, {Zafar}, {M{\o}ller}, {Ledoux}, {Kr{\"u}hler}, \&
  {Stockton}}]{Krogager16a}
{Krogager}, J.-K., {Fynbo}, J.~P.~U., {Noterdaeme}, P., {et~al.}
  2016{\natexlab{b}}, \mnras, 455, 2698

\bibitem[{{Krogager} {et~al.}(2015){Krogager}, {Geier}, {Fynbo}, {Venemans},
  {Ledoux}, {M{\o}ller}, {Noterdaeme}, {Vestergaard}, {Kangas}, {Pursimo},
  {Saturni}, \& {Smirnova}}]{Krogager15}
{Krogager}, J.-K., {Geier}, S., {Fynbo}, J.~P.~U., {et~al.} 2015, \apjs, 217, 5

\bibitem[{{Kr{\"u}hler} {et~al.}(2011){Kr{\"u}hler}, {Greiner}, {Schady},
  {Savaglio}, {Afonso}, {Clemens}, {Elliott}, {Filgas}, {Gruber}, {Kann},
  {Klose}, {K{\"u}pc{\"u}-Yolda{\c s}}, {McBreen}, {Olivares}, {Pierini},
  {Rau}, {Rossi}, {Nardini}, {Nicuesa Guelbenzu}, {Sudilovsky}, \&
  {Updike}}]{Kruhler11}
{Kr{\"u}hler}, T., {Greiner}, J., {Schady}, P., {et~al.} 2011, \aap, 534, A108

\bibitem[{{Kr{\"u}hler} {et~al.}(2008){Kr{\"u}hler}, {K{\"u}pc{\"u} Yolda{\c
  s}}, {Greiner}, {Clemens}, {McBreen}, {Primak}, {Savaglio}, {Yolda{\c s}},
  {Szokoly}, \& {Klose}}]{Kruehler08}
{Kr{\"u}hler}, T., {K{\"u}pc{\"u} Yolda{\c s}}, A., {Greiner}, J., {et~al.}
  2008, \apj, 685, 376

\bibitem[{{Kr{\"u}hler} {et~al.}(2013){Kr{\"u}hler}, {Ledoux}, {Fynbo},
  {Vreeswijk}, {Schmidl}, {Malesani}, {Christensen}, {De Cia}, {Hjorth},
  {Jakobsson}, {Kann}, {Kaper}, {Vergani}, {Afonso}, {Covino}, {de Ugarte
  Postigo}, {D'Elia}, {Filgas}, {Goldoni}, {Greiner}, {Hartoog},
  {Milvang-Jensen}, {Nardini}, {Piranomonte}, {Rossi},
  {S{\'a}nchez-Ram{\'{\i}}rez}, {Schady}, {Schulze}, {Sudilovsky}, {Tanvir},
  {Tagliaferri}, {Watson}, {Wiersema}, {Wijers}, \& {Xu}}]{Kruehler13}
{Kr{\"u}hler}, T., {Ledoux}, C., {Fynbo}, J.~P.~U., {et~al.} 2013, \aap, 557,
  A18

\bibitem[{{Ledoux} {et~al.}(2015){Ledoux}, {Noterdaeme}, {Petitjean}, \&
  {Srianand}}]{Ledoux15}
{Ledoux}, C., {Noterdaeme}, P., {Petitjean}, P., \& {Srianand}, R. 2015, \aap,
  580, A8

\bibitem[{{Ledoux} {et~al.}(2003){Ledoux}, {Petitjean}, \&
  {Srianand}}]{Ledoux03}
{Ledoux}, C., {Petitjean}, P., \& {Srianand}, R. 2003, \mnras, 346, 209

\bibitem[{{Ledoux} {et~al.}(2009){Ledoux}, {Vreeswijk}, {Smette}, {Fox},
  {Petitjean}, {Ellison}, {Fynbo}, \& {Savaglio}}]{Ledoux09}
{Ledoux}, C., {Vreeswijk}, P.~M., {Smette}, A., {et~al.} 2009, \aap, 506, 661

\bibitem[{{Ma} {et~al.}(2018){Ma}, {Ge}, {Prochaska}, {Zhang}, {Ji}, {Zhao},
  {Zhou}, {Lu}, \& {Schneider}}]{Ma18}
{Ma}, J., {Ge}, J., {Prochaska}, J.~X., {et~al.} 2018, \mnras, 474, 4870

\bibitem[{{Ma} {et~al.}(2017){Ma}, {Ge}, {Zhao}, {Prochaska}, {Zhang}, {Ji}, \&
  {Schneider}}]{Ma17}
{Ma}, J., {Ge}, J., {Zhao}, Y., {et~al.} 2017, \mnras, 472, 2196

\bibitem[{{Meegan} {et~al.}(2009){Meegan}, {Lichti}, {Bhat}, {Bissaldi},
  {Briggs}, {Connaughton}, {Diehl}, {Fishman}, {Greiner}, {Hoover}, {van der
  Horst}, {von Kienlin}, {Kippen}, {Kouveliotou}, {McBreen}, {Paciesas},
  {Preece}, {Steinle}, {Wallace}, {Wilson}, \& {Wilson-Hodge}}]{Meegan09}
{Meegan}, C., {Lichti}, G., {Bhat}, P.~N., {et~al.} 2009, \apj, 702, 791

\bibitem[{{Noterdaeme} {et~al.}(2017){Noterdaeme}, {Krogager}, {Balashev},
  {Ge}, {Gupta}, {Kr{\"u}hler}, {Ledoux}, {Murphy}, {P{\^a}ris}, {Petitjean},
  {Rahmani}, {Srianand}, \& {Ubachs}}]{Noterdaeme17}
{Noterdaeme}, P., {Krogager}, J.-K., {Balashev}, S., {et~al.} 2017, \aap, 597,
  A82

\bibitem[{{Noterdaeme} {et~al.}(2008){Noterdaeme}, {Ledoux}, {Petitjean}, \&
  {Srianand}}]{Noterdaeme08}
{Noterdaeme}, P., {Ledoux}, C., {Petitjean}, P., \& {Srianand}, R. 2008, \aap,
  481, 327

\bibitem[{{Noterdaeme} {et~al.}(2018){Noterdaeme}, {Ledoux}, {Zou},
  {Petitjean}, {Srianand}, {Balashev}, \& {L{\'o}pez}}]{Noterdaeme18}
{Noterdaeme}, P., {Ledoux}, C., {Zou}, S., {et~al.} 2018, \aap, 612, A58

\bibitem[{{Noterdaeme} {et~al.}(2015){Noterdaeme}, {Petitjean}, \&
  {Srianand}}]{Noterdaeme15}
{Noterdaeme}, P., {Petitjean}, P., \& {Srianand}, R. 2015, \aap, 578, L5

\bibitem[{{Perley} {et~al.}(2008){Perley}, {Bloom}, {Butler}, {Pollack},
  {Holtzman}, {Blake}, {Kocevski}, {Vestrand}, {Li}, {Foley}, {Bellm}, {Chen},
  {Prochaska}, {Starr}, {Filippenko}, {Falco}, {Szentgyorgyi}, {Wren},
  {Wozniak}, {White}, \& {Pergande}}]{Perley08}
{Perley}, D.~A., {Bloom}, J.~S., {Butler}, N.~R., {et~al.} 2008, \apj, 672, 449

\bibitem[{{Perley} {et~al.}(2009){Perley}, {Cenko}, {Bloom}, {Chen}, {Butler},
  {Kocevski}, {Prochaska}, {Brodwin}, {Glazebrook}, {Kasliwal}, {Kulkarni},
  {Lopez}, {Ofek}, {Pettini}, {Soderberg}, \& {Starr}}]{Perley09}
{Perley}, D.~A., {Cenko}, S.~B., {Bloom}, J.~S., {et~al.} 2009, \aj, 138, 1690

\bibitem[{{Perley} {et~al.}(2013){Perley}, {Levan}, {Tanvir}, {Cenko}, {Bloom},
  {Hjorth}, {Kr{\"u}hler}, {Filippenko}, {Fruchter}, {Fynbo}, {Jakobsson},
  {Kalirai}, {Milvang-Jensen}, {Morgan}, {Prochaska}, \&
  {Silverman}}]{Perley13}
{Perley}, D.~A., {Levan}, A.~J., {Tanvir}, N.~R., {et~al.} 2013, \apj, 778, 128

\bibitem[{{Perley} {et~al.}(2011){Perley}, {Morgan}, {Updike}, {Yuan},
  {Akerlof}, {Miller}, {Bloom}, {Cenko}, {Li}, {Filippenko}, {Prochaska},
  {Kann}, {Tanvir}, {Levan}, {Butler}, {Christian}, {Hartmann}, {Milne},
  {Rykoff}, {Rujopakarn}, {Wheeler}, \& {Williams}}]{Perley11}
{Perley}, D.~A., {Morgan}, A.~N., {Updike}, A., {et~al.} 2011, \aj, 141, 36

\bibitem[{{Petitjean} {et~al.}(2006){Petitjean}, {Ledoux}, {Noterdaeme}, \&
  {Srianand}}]{Petitjean06}
{Petitjean}, P., {Ledoux}, C., {Noterdaeme}, P., \& {Srianand}, R. 2006, \aap,
  456, L9

\bibitem[{{Planck Collaboration} {et~al.}(2016){Planck Collaboration}, {Ade},
  {Aghanim}, {Arnaud}, {Ashdown}, {Aumont}, {Baccigalupi}, {Banday},
  {Barreiro}, {Bartlett}, \& et~al.}]{Planck16}
{Planck Collaboration}, {Ade}, P.~A.~R., {Aghanim}, N., {et~al.} 2016, \aap,
  594, A13

\bibitem[{{Pontzen} \& {Pettini}(2009)}]{Pontzen09}
{Pontzen}, A. \& {Pettini}, M. 2009, \mnras, 393, 557

\bibitem[{{Prochaska} {et~al.}(2006){Prochaska}, {Chen}, \&
  {Bloom}}]{Prochaska06}
{Prochaska}, J.~X., {Chen}, H.-W., \& {Bloom}, J.~S. 2006, \apj, 648, 95

\bibitem[{{Prochaska} {et~al.}(2009){Prochaska}, {Sheffer}, {Perley}, {Bloom},
  {Lopez}, {Dessauges-Zavadsky}, {Chen}, {Filippenko}, {Ganeshalingam}, {Li},
  {Miller}, \& {Starr}}]{Prochaska09}
{Prochaska}, J.~X., {Sheffer}, Y., {Perley}, D.~A., {et~al.} 2009, \apjl, 691,
  L27

\bibitem[{{Robertson} \& {Ellis}(2012)}]{Robertson12}
{Robertson}, B.~E. \& {Ellis}, R.~S. 2012, \apj, 744, 95

\bibitem[{{Salvaterra} {et~al.}(2009){Salvaterra}, {Della Valle}, {Campana},
  {Chincarini}, {Covino}, {D'Avanzo}, {Fern{\'a}ndez-Soto}, {Guidorzi},
  {Mannucci}, {Margutti}, {Th{\"o}ne}, {Antonelli}, {Barthelmy}, {de Pasquale},
  {D'Elia}, {Fiore}, {Fugazza}, {Hunt}, {Maiorano}, {Marinoni}, {Marshall},
  {Molinari}, {Nousek}, {Pian}, {Racusin}, {Stella}, {Amati}, {Andreuzzi},
  {Cusumano}, {Fenimore}, {Ferrero}, {Giommi}, {Guetta}, {Holland}, {Hurley},
  {Israel}, {Mao}, {Markwardt}, {Masetti}, {Pagani}, {Palazzi}, {Palmer},
  {Piranomonte}, {Tagliaferri}, \& {Testa}}]{Salvaterra09}
{Salvaterra}, R., {Della Valle}, M., {Campana}, S., {et~al.} 2009, \nat, 461,
  1258

\bibitem[{{Sari} {et~al.}(1998){Sari}, {Piran}, \& {Narayan}}]{Sari98}
{Sari}, R., {Piran}, T., \& {Narayan}, R. 1998, \apjl, 497, L17

\bibitem[{{Savaglio} \& {Fall}(2004)}]{Savaglio04}
{Savaglio}, S. \& {Fall}, S.~M. 2004, \apj, 614, 293

\bibitem[{{Schady} {et~al.}(2012){Schady}, {Dwelly}, {Page}, {Kr{\"u}hler},
  {Greiner}, {Oates}, {de Pasquale}, {Nardini}, {Roming}, {Rossi}, \&
  {Still}}]{Schady12}
{Schady}, P., {Dwelly}, T., {Page}, M.~J., {et~al.} 2012, \aap, 537, A15

\bibitem[{{Schlafly} \& {Finkbeiner}(2011)}]{Schlafly11}
{Schlafly}, E.~F. \& {Finkbeiner}, D.~P. 2011, \apj, 737, 103

\bibitem[{{Schlegel} {et~al.}(1998){Schlegel}, {Finkbeiner}, \&
  {Davis}}]{Schlegel98}
{Schlegel}, D.~J., {Finkbeiner}, D.~P., \& {Davis}, M. 1998, \apj, 500, 525

\bibitem[{{Selsing} {et~al.}(2018){Selsing}, {Malesani}, {Goldoni}, {Fynbo},
  {Kr{\"u}hler}, {Antonelli}, {Arabsalmani}, {Bolmer}, {Cano}, {Christensen},
  {Covino}, {D'Avanzo}, {D'Elia}, {De Cia}, {de Ugarte Postigo}, {Flores},
  {Friis}, {Gomboc}, {Greiner}, {Groot}, {Hammer}, {Hartoog}, {Heintz},
  {Hjorth}, {Jakobsson}, {Japelj}, {Kann}, {Kaper}, {Ledoux}, {Leloudas},
  {Levan}, {Maiorano}, {Melandri}, {Milvang-Jensen}, {Palazzi}, {Palmerio},
  {Perley}, {Pian}, {Piranomonte}, {Pugliese}, {S{\'a}nchez-Ram{\'{\i}}rez},
  {Savaglio}, {Schady}, {Schulze}, {Sollerman}, {Sparre}, {Tagliaferri},
  {Tanvir}, {Th{\"o}ne}, {Vergani}, {Vreeswijk}, {Watson}, {Wiersema},
  {Wijers}, {Xu}, \& {Zafar}}]{Selsing18}
{Selsing}, J., {Malesani}, D., {Goldoni}, P., {et~al.} 2018, ArXiv (1802.07727)

\bibitem[{{Snow} \& {McCall}(2006)}]{Snow06}
{Snow}, T.~P. \& {McCall}, B.~J. 2006, \araa, 44, 367

\bibitem[{{Srianand} {et~al.}(2005){Srianand}, {Petitjean}, {Ledoux},
  {Ferland}, \& {Shaw}}]{Srianand05}
{Srianand}, R., {Petitjean}, P., {Ledoux}, C., {Ferland}, G., \& {Shaw}, G.
  2005, \mnras, 362, 549

\bibitem[{{Tanvir} {et~al.}(2009){Tanvir}, {Fox}, {Levan}, {Berger},
  {Wiersema}, {Fynbo}, {Cucchiara}, {Kr{\"u}hler}, {Gehrels}, {Bloom},
  {Greiner}, {Evans}, {Rol}, {Olivares}, {Hjorth}, {Jakobsson}, {Farihi},
  {Willingale}, {Starling}, {Cenko}, {Perley}, {Maund}, {Duke}, {Wijers},
  {Adamson}, {Allan}, {Bremer}, {Burrows}, {Castro-Tirado}, {Cavanagh}, {de
  Ugarte Postigo}, {Dopita}, {Fatkhullin}, {Fruchter}, {Foley}, {Gorosabel},
  {Kennea}, {Kerr}, {Klose}, {Krimm}, {Komarova}, {Kulkarni}, {Moskvitin},
  {Mundell}, {Naylor}, {Page}, {Penprase}, {Perri}, {Podsiadlowski}, {Roth},
  {Rutledge}, {Sakamoto}, {Schady}, {Schmidt}, {Soderberg}, {Sollerman},
  {Stephens}, {Stratta}, {Ukwatta}, {Watson}, {Westra}, {Wold}, \&
  {Wolf}}]{Tanvir09}
{Tanvir}, N.~R., {Fox}, D.~B., {Levan}, A.~J., {et~al.} 2009, \nat, 461, 1254

\bibitem[{{Tanvir} {et~al.}(2018){Tanvir}, {Laskar}, {Levan}, {Perley}, {Zabl},
  {Fynbo}, {Rhoads}, {Cenko}, {Greiner}, {Wiersema}, {Hjorth}, {Cucchiara},
  {Berger}, {Bremer}, {Cano}, {Cobb}, {Covino}, {D'Elia}, {Fong}, {Fruchter},
  {Goldoni}, {Hammer}, {Heintz}, {Jakobsson}, {Kann}, {Kaper}, {Klose},
  {Knust}, {Kr{\"u}hler}, {Malesani}, {Misra}, {Nicuesa Guelbenzu}, {Pugliese},
  {S{\'a}nchez-Ram{\'{\i}}rez}, {Schulze}, {Stanway}, {de Ugarte Postigo},
  {Watson}, {Wijers}, \& {Xu}}]{Tanvir18}
{Tanvir}, N.~R., {Laskar}, T., {Levan}, A.~J., {et~al.} 2018, \apj, 865, 107

\bibitem[{{Tumlinson} {et~al.}(2007){Tumlinson}, {Prochaska}, {Chen},
  {Dessauges-Zavadsky}, \& {Bloom}}]{Tumlinson07}
{Tumlinson}, J., {Prochaska}, J.~X., {Chen}, H.-W., {Dessauges-Zavadsky}, M.,
  \& {Bloom}, J.~S. 2007, \apj, 668, 667

\bibitem[{{van der Horst} {et~al.}(2009){van der Horst}, {Kouveliotou},
  {Gehrels}, {Rol}, {Wijers}, {Cannizzo}, {Racusin}, \&
  {Burrows}}]{VanDerHorst09}
{van der Horst}, A.~J., {Kouveliotou}, C., {Gehrels}, N., {et~al.} 2009, \apj,
  699, 1087

\bibitem[{{Vernet} {et~al.}(2011){Vernet}, {Dekker}, {D'Odorico}, {Kaper},
  {Kjaergaard}, {Hammer}, {Randich}, {Zerbi}, {Groot}, {Hjorth}, {Guinouard},
  {Navarro}, {Adolfse}, {Albers}, {Amans}, {Andersen}, {Andersen}, {Binetruy},
  {Bristow}, {Castillo}, {Chemla}, {Christensen}, {Conconi}, {Conzelmann},
  {Dam}, {de Caprio}, {de Ugarte Postigo}, {Delabre}, {di Marcantonio},
  {Downing}, {Elswijk}, {Finger}, {Fischer}, {Flores}, {Fran{\c c}ois},
  {Goldoni}, {Guglielmi}, {Haigron}, {Hanenburg}, {Hendriks}, {Horrobin},
  {Horville}, {Jessen}, {Kerber}, {Kern}, {Kiekebusch}, {Kleszcz}, {Klougart},
  {Kragt}, {Larsen}, {Lizon}, {Lucuix}, {Mainieri}, {Manuputy}, {Martayan},
  {Mason}, {Mazzoleni}, {Michaelsen}, {Modigliani}, {Moehler}, {M{\o}ller},
  {Norup S{\o}rensen}, {N{\o}rregaard}, {P{\'e}roux}, {Patat}, {Pena}, {Pragt},
  {Reinero}, {Rigal}, {Riva}, {Roelfsema}, {Royer}, {Sacco}, {Santin},
  {Schoenmaker}, {Spano}, {Sweers}, {Ter Horst}, {Tintori}, {Tromp}, {van
  Dael}, {van der Vliet}, {Venema}, {Vidali}, {Vinther}, {Vola}, {Winters},
  {Wistisen}, {Wulterkens}, \& {Zacchei}}]{Vernet11}
{Vernet}, J., {Dekker}, H., {D'Odorico}, S., {et~al.} 2011, \aap, 536, A105

\bibitem[{{Vladilo} {et~al.}(2006){Vladilo}, {Centuri{\'o}n}, {Levshakov},
  {P{\'e}roux}, {Khare}, {Kulkarni}, \& {York}}]{Vladilo06}
{Vladilo}, G., {Centuri{\'o}n}, M., {Levshakov}, S.~A., {et~al.} 2006, \aap,
  454, 151

\bibitem[{{Vladilo} \& {P{\'e}roux}(2005)}]{Vladilo05}
{Vladilo}, G. \& {P{\'e}roux}, C. 2005, \aap, 444, 461

\bibitem[{{Vreeswijk} {et~al.}(2007){Vreeswijk}, {Ledoux}, {Smette}, {Ellison},
  {Jaunsen}, {Andersen}, {Fruchter}, {Fynbo}, {Hjorth}, {Kaufer}, {M{\o}ller},
  {Petitjean}, {Savaglio}, \& {Wijers}}]{Vreeswijk07}
{Vreeswijk}, P.~M., {Ledoux}, C., {Smette}, A., {et~al.} 2007, \aap, 468, 83

\bibitem[{{Vreeswijk} {et~al.}(2011){Vreeswijk}, {Ledoux}, {Smette}, {Ellison},
  {Jaunsen}, {Andersen}, {Fruchter}, {Fynbo}, {Hjorth}, {Kaufer}, {M{\o}ller},
  {Petitjean}, {Savaglio}, \& {Wijers}}]{Vreeswijk11}
{Vreeswijk}, P.~M., {Ledoux}, C., {Smette}, A., {et~al.} 2011, \aap, 532, C3

\bibitem[{{Watson} {et~al.}(2006){Watson}, {Fynbo}, {Ledoux}, {Vreeswijk},
  {Hjorth}, {Smette}, {Andersen}, {Aoki}, {Augusteijn}, {Beardmore}, {Bersier},
  {Castro Cer{\'o}n}, {D'Avanzo}, {Diaz-Fraile}, {Gorosabel}, {Hirst},
  {Jakobsson}, {Jensen}, {Kawai}, {Kosugi}, {Laursen}, {Levan}, {Masegosa},
  {N{\"a}r{\"a}nen}, {Page}, {Pedersen}, {Pozanenko}, {Reeves}, {Rumyantsev},
  {Shahbaz}, {Sharapov}, {Sollerman}, {Starling}, {Tanvir}, {Torstensson}, \&
  {Wiersema}}]{Watson06}
{Watson}, D., {Fynbo}, J.~P.~U., {Ledoux}, C., {et~al.} 2006, \apj, 652, 1011

\bibitem[{{Watson} \& {Jakobsson}(2012)}]{Watson12}
{Watson}, D. \& {Jakobsson}, P. 2012, \apj, 754, 89

\bibitem[{{Whalen} {et~al.}(2008){Whalen}, {Prochaska}, {Heger}, \&
  {Tumlinson}}]{Whalen08}
{Whalen}, D., {Prochaska}, J.~X., {Heger}, A., \& {Tumlinson}, J. 2008, \apj,
  682, 1114

\bibitem[{{Wijers} {et~al.}(1998){Wijers}, {Bloom}, {Bagla}, \&
  {Natarajan}}]{Wijers98}
{Wijers}, R.~A.~M.~J., {Bloom}, J.~S., {Bagla}, J.~S., \& {Natarajan}, P. 1998,
  \mnras, 294, L13

\bibitem[{{Winkler} {et~al.}(2003){Winkler}, {Courvoisier}, {Di Cocco},
  {Gehrels}, {Gim{\'e}nez}, {Grebenev}, {Hermsen}, {Mas-Hesse}, {Lebrun},
  {Lund}, {Palumbo}, {Paul}, {Roques}, {Schnopper}, {Sch{\"o}nfelder},
  {Sunyaev}, {Teegarden}, {Ubertini}, {Vedrenne}, \& {Dean}}]{Winkler03}
{Winkler}, C., {Courvoisier}, T.~J.-L., {Di Cocco}, G., {et~al.} 2003, \aap,
  411, L1

\bibitem[{{Wiseman} {et~al.}(2017){Wiseman}, {Schady}, {Bolmer}, {Kr{\"u}hler},
  {Yates}, {Greiner}, \& {Fynbo}}]{Wiseman17}
{Wiseman}, P., {Schady}, P., {Bolmer}, J., {et~al.} 2017, \aap, 599, A24

\bibitem[{{Woosley} \& {Bloom}(2006)}]{Woosley06}
{Woosley}, S.~E. \& {Bloom}, J.~S. 2006, \araa, 44, 507

\bibitem[{{Zafar} {et~al.}(2018{\natexlab{a}}){Zafar}, {Heintz}, {Fynbo},
  {Malesani}, {Bolmer}, {Ledoux}, {Arabsalmani}, {Kaper}, {Campana},
  {Starling}, {Selsing}, {Kann}, {de Ugarte Postigo}, {Schweyer},
  {Christensen}, {M{\o}ller}, {Japelj}, {Perley}, {Tanvir}, {D'Avanzo},
  {Hartmann}, {Hjorth}, {Covino}, {Sbarufatti}, {Jakobsson}, {Izzo},
  {Salvaterra}, {D'Elia}, \& {Xu}}]{Zafar18b}
{Zafar}, T., {Heintz}, K.~E., {Fynbo}, J.~P.~U., {et~al.} 2018{\natexlab{a}},
  \apjl, 860, L21

\bibitem[{{Zafar} {et~al.}(2018{\natexlab{b}}){Zafar}, {M{\o}ller}, {Watson},
  {Lattanzio}, {Hopkins}, {Karakas}, {Fynbo}, {Tanvir}, {Selsing}, {Jakobsson},
  {Heintz}, {Kann}, {Groves}, {Kulkarni}, {Covino}, {D'Elia}, {Japelj},
  {Corre}, \& {Vergani}}]{Zafar18c}
{Zafar}, T., {M{\o}ller}, P., {Watson}, D., {et~al.} 2018{\natexlab{b}},
  \mnras, 480, 108

\bibitem[{{Zafar} \& {Watson}(2013)}]{Zafar13}
{Zafar}, T. \& {Watson}, D. 2013, \aap, 560, A26

\bibitem[{{Zafar} {et~al.}(2012){Zafar}, {Watson}, {El{\'{\i}}asd{\'o}ttir},
  {Fynbo}, {Kr{\"u}hler}, {Schady}, {Leloudas}, {Jakobsson}, {Th{\"o}ne},
  {Perley}, {Morgan}, {Bloom}, \& {Greiner}}]{Zafar12}
{Zafar}, T., {Watson}, D., {El{\'{\i}}asd{\'o}ttir}, {\'A}., {et~al.} 2012,
  \apj, 753, 82

\bibitem[{{Zafar} {et~al.}(2011{\natexlab{a}}){Zafar}, {Watson}, {Fynbo},
  {Malesani}, {Jakobsson}, \& {de Ugarte Postigo}}]{Zafar11a}
{Zafar}, T., {Watson}, D., {Fynbo}, J.~P.~U., {et~al.} 2011{\natexlab{a}},
  \aap, 532, A143

\bibitem[{{Zafar} {et~al.}(2018{\natexlab{c}}){Zafar}, {Watson}, {M{\o}ller},
  {Selsing}, {Fynbo}, {Schady}, {Wiersema}, {Levan}, {Heintz}, {de Ugarte
  Postigo}, {D'Elia}, {Jakobsson}, {Bolmer}, {Japelj}, {Covino}, {Gomboc}, \&
  {Cano}}]{Zafar18a}
{Zafar}, T., {Watson}, D., {M{\o}ller}, P., {et~al.} 2018{\natexlab{c}},
  \mnras, 479, 1542

\bibitem[{{Zafar} {et~al.}(2011{\natexlab{b}}){Zafar}, {Watson}, {Tanvir},
  {Fynbo}, {Starling}, \& {Levan}}]{Zafar11b}
{Zafar}, T., {Watson}, D.~J., {Tanvir}, N.~R., {et~al.} 2011{\natexlab{b}},
  \apj, 735, 2

\bibitem[{{Zou} {et~al.}(2018){Zou}, {Petitjean}, {Noterdaeme}, {Ledoux},
  {Krogager}, {Fathivavsari}, {Srianand}, \& {L{\'o}pez}}]{Zou18}
{Zou}, S., {Petitjean}, P., {Noterdaeme}, P., {et~al.} 2018, \aap, 616, A158

\end{thebibliography}

\end{document}